\documentclass[12pt, oneside]{article}   	
\usepackage{geometry}                		
\usepackage{graphicx}				
\usepackage{amssymb,amsmath,bbm,bm,pdflscape,amsthm}
\usepackage{setspace,booktabs,threeparttable,float}
\allowdisplaybreaks

\usepackage[longnamesfirst]{natbib}
\usepackage[noblocks]{authblk}

\newtheorem{theorem}{Theorem}[section]

\newtheorem{condition}{Condition}

\newtheorem{assumption}{Assumption}
\newtheorem{proposition}{Proposition}[section]
\newtheorem{definition}{Definition}[section]
\newtheorem{corollary}{Corollary}[section]
\newtheorem{lemma}[theorem]{Lemma}
\newtheorem{remark}{Remark}

\newcommand{\norm}[1]{\left\lVert#1\right\rVert}

\newcommand{\indep}{\perp \!\!\! \perp}

\newcommand\numberthis{\addtocounter{equation}{1}\tag{\theequation}}

\newtheorem*{subassumption*}{\assumptionnumber}
\providecommand{\assumptionnumber}{}


\newtheorem{assumptionalt}{Assumption}[assumption]

\newenvironment{assumptionp}[1]{
	\renewcommand\theassumptionalt{#1}
	\assumptionalt
}{\endassumptionalt}

\title{Difference-in-Differences with Interference}
\author{Ruonan Xu\thanks{Department of Economics, Rutgers University, ruonan.xu@rutgers.edu}}
\date{}	

\begin{document}
\maketitle

\begin{abstract}

In many scenarios, such as the evaluation of place-based policies, potential outcomes are not only dependent upon the unit's own treatment but also its neighbors' treatment. Despite this, ``difference-in-differences” (DID) type estimators typically ignore such interference among neighbors. I show in this paper that the canonical DID estimators generally fail to identify interesting causal effects in the presence of neighborhood interference. To incorporate interference structure into DID estimation, I propose doubly robust estimators for the direct average treatment effect on the treated as well as the average spillover effects under a modified parallel trends assumption. I later relax common restrictions in the literature, such as immediate neighborhood interference and correctly specified spillover functions. Moreover, robust inference is discussed based on the asymptotic distribution of the proposed estimators. 


\end{abstract}

\section{Introduction}
According to the stable unit treatment value assumption (SUTVA), potential outcomes only depend on one's own treatment assignment. In many cases, SUTVA fails due to an unknown interference structure among neighbors. In the fields of environmental economics, urban economics, labor economics, criminal justice, and many other fields of social sciences, place-based policies often generate spillover effects. One example is minimum wage increase in Seattle studied by \cite{jardim2022boundary}. Through the channels of competition in the regional labor market for workers and the possibility of relocation of businesses, they find that significant spillover effects on wages and hours are seen up to a 40-minute drive from Seattle city limits. 

When spillover effects are of interest, one often needs to observe the entire population. For example, we can typically collect information about all counties in the United States. In the example above, \cite{jardim2022boundary} use administrative employment records in the state of Washington. In this way, we do not need to deal with partial observation of other units' treatment status and can allow for diverse interference patterns based on access to the entire treatment assignment vector. If we take sampling from the superpopulation/infinite population approach literally, what we are estimating turns out to be the spillover effect in a researcher's sample with missing neighbors unless interactions are restricted within clusters of friends or household members and one randomly samples clusters.\footnote{See further explanation below equation (7) on page 537 in \cite{manski1993identification}. } 

In this paper, I study the ``difference-in-differences" (DID) type estimators that allow interference from a finite population perspective, where inference is conditional on covariates and the whole population is observed. This approach is closest to the conditional inference discussed by \cite{abadie2014inference} and \cite{jin2023tailored}. Conditional treatment effect parameters have also been mentioned in \cite{abadie2002simple}, \cite{imbens2004nonparametric}, and \cite{balzer2015targeted}. Recently, \cite{viviano2024policy} adopts the same inference framework when studying optimal treatment allocation under network interference. 

The conditional inference approach adopted here allows arbitrary spatial correlation and nonstationarity of covariates. Meanwhile, stochastic potential outcomes allow the possibility of modeling the conditional means of the outcome variables and incorporate more uncertainty into the inference framework. Consequently, the proposed estimators below are more robust to model specification with a straightforward causal interpretation. 
One could legitimately argue that researchers should not stick to a single inference framework. That said, many attribute variables containing locational information and neighborhood characteristics, such as landlocked status, are deemed non-stochastic for spatial data. I therefore consider the current approach a natural starting point for studying population interference/spillover effects.\footnote{The conditional inference framework only affects the definition of the parameter of interest and the inference later on. Estimation is not affected by whether the current conditional inference, design-based inference, or superpopulation framework is adopted.} 

One challenge of incorporating interference is the modeling of spillover functions. Following the interference literature I introduce exposure mapping, which is a function that maps an assignment vector to an exposure value \citep{aronow2017estimating, manski2013identification}. One leading example is specifying exposure as the average value of treatment statuses of neighbors within a distance $\bar{\rho}$ of a unit $i$. I start with the case where exposure mapping is assumed to be known and correctly specified. Later on, I generalize the analysis to the case where exposure mapping could be misspecified. Regardless of the correct specification of the exposure mapping, the same estimator and inference procedure is proposed for the direct average treatment effect on the treated (DATT) and the spillover effect defined in the current paper. As a result, practitioners can still use the spillover function/exposure mapping they choose based on their domain/institutional knowledge. Inspired by \cite{savje2023causal}, the estimand remains well-defined under misspecification of the exposure. Only its interpretation needs to be modified. In addition, the assignment variables are allowed to be spatially correlated as is often the case in practice with spatial data. 

Putting all the pieces together, I propose doubly robust estimators for the direct treatment effect and spillover effect. The proposed doubly robust estimator is a modified version of the augmented inverse probability weighting (AIPW) estimator, which only requires correct specification of either the propensity scores of treatment/exposure or the conditional mean of the outcomes. The conditional inference approach in the current paper leads to a different variance-covariance matrix which may require a new variance estimator when necessary. 

Besides the main contribution above, I study the identification of canonical DID estimators available in the literature in Section \ref{sec:canonic} below. I provide conditions under which canonical estimators can still identify meaningful causal parameters. This discussion alone would be of interest to practitioners. The proposed doubly robust estimators are applied in Section \ref{sec:application} to study the policy effect of special economic zones (SEZ) in China. Appendix \ref{sec:guide} summarizes the detailed steps for direct effect estimation. 

\textbf{Related Literature}: \cite{delgado2015difference} and \cite{butts2021difference} allow interference in DID estimation in a two-way fixed effects (TWFE) estimating equation (often without covariates) from a superpopulation perspective. \cite{huber2021framework} also propose a DID approach to estimate spillover effect and total effect within a superpopulation framework.\footnote{Their potential outcomes are defined as functions of individual and regional treatments, where individual treatment status is a function of the regional treatment. Therefore, \cite{huber2021framework} is more applicable to studies of local equilibrium effects.} I instead propose doubly robust estimators with flexible choice of exposure mappings and adopt a finite population framework. In addition, all three papers mentioned above share some or all the restrictions of the general interference literature. More specifically, most methodological literature studies spillover effects in a single cross section of experimental data and assumes partial interference or limits interference to immediate neighbors. Additionally, they assume that the function of dependence on neighbors' treatments is known and correctly specified. See, for instance, \cite{hudgens2008toward} and \cite{aronow2017estimating}. I relax these assumptions in a DID context in Section \ref{misspecify} below discussing misspecification of the exposure mapping, adapting tools from \cite{savje2023causal} and \cite{leung2022causal}. Design-based DID estimation has been studied by \cite{athey2022design} and \cite{rambachan2020design}, but they keep the SUTVA. \cite{sant2020doubly} have proposed AIPW estimators in the DID context, maintaining SUTVA and the superpopulation perspective.

\section{Setup}
\subsection{Environment}\label{setup}
I start with the relatively simple setting of panel data with two time periods; $t=1,2$ stands for the time period before and after treatment respectively. Consider a sequence of lattices of (possibly) unevenly placed locations in $\mathbb{R}^d$, $\{D_M\subseteq \mathbb{R}^d, d\geq 1\}$, where $M$ indexes the sequence of finite populations. In a finite population setting, the parameters of interest are defined within the finite population. Because I consider the case where the sample coincides with the population for spatial data, I let the population size $|D_M|$ diverge to infinity in deriving the asymptotic properties of the proposed estimators and conducting inference, where $|V|$ denotes the cardinality of a finite subset $V\subseteq D_M$. 

I briefly summarize the notation used throughout the paper. I adopt the metric $\rho(i,j)=\max_{1\leq l\leq d}|j_l-i_l|$ in space $\mathbb{R}^d$, where $i_l$ is the $l$-th component of $i$. This metric could capture geographic distance or some economic distance between units $i$ and $j$. The distance between any subsets $K,V\subseteq D_M$ is defined as $\rho(K,V)=\inf\{\rho(i,j): i\in K\mbox{ and } j\in V\}$. For any random vector $X$, $\norm{X}_p=\big(\mathbbm{E}\norm{X}^p\big)^{1/p}$, $p\geq 1$, denotes its $L_p$-norm. Lastly, $C$ denotes a generic positive constant that may vary under different circumstances.

For each unit $i$ in the population, there is a stochastic assignment variable $W_i\in \{0, 1\}$, a vector of fixed attributes $z_i=(z_i^{ind},z_i^{neigh})$ that possibly includes attributes of $i$'s neighborhood $z_i^{neigh}$ in addition to individual characteristics $z_i^{ind}$, and a vector of stochastic unobservables $U_{it}$. The potential outcome function for any $i\in D_M$ is defined as $h_{it}(\cdot): \{0,1\}^{|D_M|}\times \mathbb{R}^{\dim(z_i)}\times \mathbb{R}^{\dim(U_{it})} \to \mathbb{R}$. I emphasize the treatment vector of the entire population by denoting the potential outcomes as $y_{it}(w_i,\bm{w}_{-i})=h(w_i,\bm{w}_{-i}, z_{i}, U_{it})$, where $\bm{w}_{-i}=\{w_j, j\in D_M, j\neq i\}$.\footnote{As in \cite{manski2013identification}, the potential outcome function defined here can be considered as the response function, namely the reduced form of structural equations where the structural potential outcome may depend on other units' treatments as well as outcomes.} The dependence of the potential outcomes on the fixed attributes and stochastic unobservables is indicated by its $i,t$ subscript. The realized potential outcomes are denoted by $Y_{it}=y_{it}(\bm{W})$. Notice that $(\bm{W}, \bm{z}, \bm{Y},\bm{U})=\{(W_i, z_i, Y_{it}(\cdot),U_{it}), i\in D_M, M\geq 1\}$ are triangular arrays of random fields defined on a probability space $(\Omega, \mathcal{F}, P)$. Exposure mapping is defined by the function $G_i=G(i, \bm{W}_{-i})\in \mathcal{G}$, where $\mathcal{G}$ is a discrete set.\footnote{Given our finite population setting and a binary individual treatment variable $W_i$, $\mathcal{G}$ is a finite set. This is also the approach taken by \cite{aronow2017estimating} and \cite{leung2022causal}.} Therefore, $G(\cdot)$ maps the treatment status of all units except $i$ to an exposure value.  

In line with common practices, empirical researchers can construct the exposure mapping $G(\cdot)$ in the following manner, which is assumed to capture the true interference structure. Given a fixed $K$, define the $K$-neighborhood of the unit $i$ as 
\[
\mathcal{N}(i,K)=\{j\in D_M: \rho(i,j)\leq K, j\neq i \}
\]
Let $\bm{w}_{\mathcal{N}(i,K)}=(w_j: j\in \mathcal{N}(i,K))$ be the treatment vector of units within $i$'s $K$-neighborhood. There exists $K<\infty$ such that for all $\bm{w}_{-i}$ and $\bm{w}'_{-i}$ such that $\bm{w}_{\mathcal{N}(i,K)}=\bm{w}'_{\mathcal{N}(i,K)}$, $G(i,\bm{w}_{-i})=G(i,\bm{w}'_{-i})$. As a result, the specified exposure mapping function restricts spillover effects within the immediate $K$-neighborhood of each unit.\footnote{With correctly specified exposure mapping, asymptotic distribution of the proposed estimators below can be derived for general exposure mapping not restricted to $K$-neighborhood interference under a different set of local dependence assumptions than the ones listed below. Later on, I relax the immediate $K$-neighborhood interference to allow for potential misspecification of the exposure mapping. With the current approach of constructing the exposure mapping, the asymptotic results can be derived in a unified framework accommodating both correctly or incorrectly specified exposure mapping.} A leading example is $G_i=\sum_{j\in D_M, j\neq i} A_{ij}W_j/\sum_{j\in D_M,j\neq i} A_{ij}$, where $A_{ij}=1$ if the distance between units $i$ and $j$ is within a cutoff $K$. Given a correctly specified exposure mapping with $G(i,\bm{w}_{-i})=g$, $\tilde{y}_{i2}(w_i,g)=y_{i2}(w_i,\bm{w}_{-i})$.

\subsection{Estimands of Interest}
This paper is interested in the expected finite population average, i.e., the average of the expected potential outcome across all units in the finite population. In other words, I focus on conditional inference given fixed attributes $z_i$; see \cite{abadie2014inference} and \cite{jin2023tailored} for detailed discussion of conditional parameters and conditional inference. 

There are two types of estimands of interest. For the main part of the paper, I will focus on the first type, the direct treatment effect. 
There is more than one way to define the parameter of interest. As an analogy to the expected average treatment effect in \cite{savje2021average}, the overall direct effect can be defined as 
\[
\tau=\frac{1}{|D_M|}\sum_{i\in D_M}\mathbbm{E}\big[y_{i2}(1,\bm{W}_{-i})-y_{i2}(0,\bm{W}_{-i})|W_i=1, z_i\big],
\]
which marginalizes over the treatment assignment vector. The overall direct effect is a natural extension of the average treatment effect on the treated (ATT) as it coincides with the ATT when units do not interfere ($y_{i2}(W_i,\bm{W}_{-i})$ reduces to $y_{i2}(W_i)$).  

Often the time, in addition to a summary of the direct effects, researchers can also be interested in direct effect at different exposure levels. The overall direct effect is highly related to the direct average treatment effect on the treated (DATT) at exposure levels $g\in \mathcal{G}$ defined in equation (\ref{eq0}) below.\footnote{Their relationship is explained in equation (\ref{eq:p1}) in Appendix \ref{sec:proof}.}
\begin{equation}
\label{eq0}
\begin{aligned}
\tau(g)=&\frac{1}{|D_M|}\sum_{i\in D_M}\mathbbm{E}\big[y_{i2}(1,\bm{W}_{-i})-y_{i2}(0,\bm{W}_{-i})|W_i=1, G_i=g, z_i\big]\\
=&\frac{1}{|D_M|}\sum_{i\in D_M}\mathbbm{E}\big[\tilde{y}_{i2}(1,g)-\tilde{y}_{i2}(0,g)|W_i=1, G_i=g, z_i\big]
\end{aligned}
\end{equation}
Without interference, $\tau(g)$ becomes a conditional ATT with $G_i=g$ serving as another characteristic of unit $i$.
I focus on the identification and estimation of DATT as the estimation of the overall direct effect follows using weighted averages when the weights are correctly specified.\footnote{As shown in Section 3.1, the canonical DID estimand, which ignores interference, fails to identify $\tau$ for spatially correlated assignments. Notice that I do not propose an estimator for the overall direct effect in this paper. \cite{savje2021average} show when common estimators, such as the Horvitz-Thompson and H\'ajek estimators, are consistent for the expected average treatment effect without proposing a new estimator.} Also, by contrasting different exposure levels, the definition of DATT facilitates the discussion of the second estimand, the spillover effect. In the interest of space, this is delegated to Appendix B. 

The all-or-nothing effect, $\frac{1}{|D_M|}\sum_{i\in D_M}\mathbbm{E}\big[y_{i2}(\bm{1})-y_{i2}(\bm{0})| z_i\big]$, where $\bm{1}$ and $\bm{0}$ are unit and zero vectors, cannot be consistently estimated (\cite{basse2018limitations}). Instead, direct effects and spillover effects summarize different aspects of policy effects. For the DATT I consider in the main text, it captures the direct treatment effect at different exposure levels. In a vaccination example, if the direct effect of vaccinating an additional individual is almost zero given that a certain fraction of the population are already vaccinated, this can serve as an indicator of herd immunity being achieved. 

I use the empirical application in Section \ref{sec:application} below to illustrate the relevant variables and estimands. The data I use come from \cite{lu2019place}, who study how China's SEZ policy impacts various outcomes $Y_{it}$, such as the logarithm of firm output at the village level. If village $i$ is located within the boundaries of a SEZ, direct treatment variable $W_i$ is equal to one and otherwise, it is equal to zero. Exposure mapping $G_i$ is a binary variable equal to one if the leave-one-out ratio of SEZ villages to the total number of villages in the county $c$ in which village $i$ is located is greater than the mean ratio among all counties. In this case, the DATT captures the direct effect of establishing a SEZ in village $i$ given the fraction of SEZ villages within a county. It is possible that the direct effect is lower when there is a higher proportion of neighboring SEZ villages. This can help determine whether establishing an additional SEZ is cost-effective.

\section{Identification}
The first question when relaxing SUTVA is what the canonical DID estimator identifies if spillover effects are incorrectly ignored. Namely, will the canonical DID estimator still consistently estimate ATT in the presence of interference? \cite{forastiere2021identification} discuss bias of the difference-in-means estimator when SUTVA is wrongly assumed in observational studies on networks. To my knowledge, the literature has not yet investigated DID type estimators.  
To facilitate the discussion of identification, I impose the following assumptions. 
\begin{assumption}\label{overlap}
(Overlap)
$\forall\ i\in D_M$, there exists $\epsilon >0$ such that $\epsilon<p(z_i)<1-\epsilon$, $\pi_{1g}(z_i)>\epsilon$, and $\pi_{0g}(z_i)>\epsilon$, where
\begin{equation}
\label{ps}
p(z_i)=P(W_i=1|z_i),
\end{equation}
\begin{equation}
\label{psg1}
\pi_{1g}(z_i)=P(G_i=g|W_i=1, z_i),
\end{equation}
and
\begin{equation}
\label{psg0}
\pi_{0g}(z_i)=P(G_i=g|W_i=0, z_i).
\end{equation}
\end{assumption}

To simplify notation, I assume that the overlap assumption applies to every unit in the population. With certain exposure mapping specifications, this might not be plausible. An easy fix is to change the estimand by averaging over the subpopulation where $G_i$ can take on the value $g$. Failure to satisfy the overlap condition for $p(z_i)$ is trickier. If one is willing to move the goalpost by redefining the population, one can drop units that always or never take treatment. The good news is that for the redefined population, we can still observe the treatment assignment vector of the original population since the treatment status of the dropped units is fixed and known. This way, dropping the always or never takers will not affect the exposure mapping. On the other hand, to deal with weak overlap conditions in practice without changing the population or estimand, one can consider approaches proposed by \cite{ma2020robust} and \cite{man2023doubly} to trim propensity scores and correct the resulting bias simultaneously. 

\begin{assumption}\label{anticipation}
(No Anticipation) 
\[
y_{i1}(w_i,\bm{w}_{-i})=y_{i1}(0,\underline{0})
\]
\end{assumption}
Assumption \ref{anticipation} requires that the potential outcome in the first time period prior to treatment is always equal to the potential outcome without treatment nor spillover ($\bm{w}_{-i}=\underline{0}$). 
The no-anticipation assumption is quite standard in the literature, sometimes implicitly assumed.

With correctly specified exposure mapping, I impose the following parallel trends assumption: 
\begin{assumption}\label{trueparallel}
(Parallel Trends)
For any $g\in \mathcal{G}$ and $\forall\ i$,
\begin{equation}\label{true}
\begin{aligned}
&\mathbbm{E}\big[\tilde{y}_{i2}(0, g)|W_i=1,G_i=g, z_i\big]-\mathbbm{E}\big[y_{i1}(0,\underline{0})|W_i=1,G_i=g,z_i\big]\\
=&\mathbbm{E}\big[\tilde{y}_{i2}(0, g)|W_i=0,G_i=g, z_i\big]-\mathbbm{E}\big[y_{i1}(0,\underline{0})|W_i=0,G_i=g,z_i\big]
\end{aligned}
\end{equation}
\end{assumption}
Notice that in the parallel trends assumption, as no one is treated at $t=1$  there is no spillover in the potential outcome function in the first time period. Namely, moving from the first to the second time period, in the absence of direct treatment the conditional mean of the potential outcomes for the treated and the untreated with the same level of exposure in the second time period follows the same trend. Equation (\ref{true}) serves as our starting point for identification.

There is a growing literature on justification and falsification of the parallel trends assumption under SUTVA; see, for instance, \cite{ghanem2022selection} and \cite{roth2023parallel}. When parallel trends might be violated, \cite{rambachan2023more} present confidence sets for the identified set of treatment effects. The extension of these analyses to parallel trends with interference is outside the scope of the current paper and left as future research. 

\subsection{Canonical DID}\label{sec:canonic}
The usual ATT under the SUTVA is 
\[
\tilde{\tau}=\frac{1}{|D_M|}\sum_{i\in D_M}\mathbbm{E}\big[y_{i2}(1)-y_{i2}(0)|W_i=1,z_i\big].
\]
Here, the potential outcomes are determined solely by unit $i$'s own treatment.
Suppose the canonical DID estimator consistently estimates
\begin{align*}
\tau_{canonic}=\frac{1}{|D_M|}\sum_{i\in D_M}\Big[\mathbbm{E}(Y_{i2}-Y_{i1}|W_i=1,z_i)-\mathbbm{E}(Y_{i2}-Y_{i1}|W_i=0,z_i)\Big].
\end{align*}
Examples include the TWFE linear estimating equation in Remark 1 in \cite{sant2020doubly} under the additional restrictions of the data generating process therein, as well as the inverse probability weighting (IPW) estimator in \cite{abadie2005semiparametric}.
If the usual (conditional) parallel trends assumption holds without interference, $\tau_{canonic}$ would be equivalent to $\tilde{\tau}$.

If SUTVA is violated, $\tilde{\tau}$ is not well defined as the potential outcome should depend on the entire assignment vector. Also, DATT is generally determined by the specified exposure level. As a result, I use the overall direct effect as a benchmark for comparison. 
Using the law of iterated expectations, $\tau$ and $\tau_{canonic}$ can be decomposed in the following way:
\begin{align*}
\tau=&\frac{1}{|D_M|}\sum_{i\in D_M}\bigg\{\sum_{g\in \mathcal{G}}\Big[\mathbbm{E}(Y_{i2}|W_i=1,G_i=g,z_i)-\mathbbm{E}(Y_{i1}|W_i=1,G_i=g,z_i)\Big]P(G_i=g|W_i=1,z_i)\\
&-\sum_{g\in \mathcal{G}}\Big[\mathbbm{E}(Y_{i2}|W_i=0,G_i=g,z_i)-\mathbbm{E}(Y_{i1}|W_i=0,G_i=g,z_i)\Big]P(G_i=g|W_i=1,z_i)\bigg\}
\end{align*}
\begin{align*}
\tau_{canonic}=&\frac{1}{|D_M|}\sum_{i\in D_M}\bigg\{\sum_{g\in \mathcal{G}}\Big[\mathbbm{E}(Y_{i2}|W_i=1,G_i=g,z_i)\\
&-\mathbbm{E}(Y_{i1}|W_i=1,G_i=g,z_i)\Big]P(G_i=g|W_i=1,z_i)\\
&-\sum_{g\in \mathcal{G}}\Big[\mathbbm{E}(Y_{i2}|W_i=0,G_i=g,z_i)-\mathbbm{E}(Y_{i1}|W_i=0,G_i=g,z_i)\Big]P(G_i=g|W_i=0,z_i)\bigg\}
\end{align*}
\begin{proposition}\label{pro1}
Under Assumptions \ref{anticipation} and \ref{trueparallel},
$\tau_{canonic}\neq \tau$ unless $P(G_i=g|W_i=0,z_i)=P(G_i=g|W_i=1,z_i)$ for any $g\in \mathcal{G}$ and $\forall i$ .
\end{proposition}

The equality of the conditional probabilities holds if $G_i\indep W_i\ |\ z_i$. However, conditional independence can be easily violated if either of the following is true: (\romannumeral 1) $G_i$ and $W_i$ are linked through covariates not included in $z_i$; (\romannumeral 2) neighbors' behavior affects unit $i$'s treatment uptake; (\romannumeral 3) similar neighborhood characteristics drive the assignment mechanism; see \cite{forastiere2021identification} for a parallel discussion allowing interference on networks under unconfoundedness. 
As a consequence, the overall direct effect can be either underestimated or overestimated by the canonical DID estimators. 

\begin{remark}\label{remark1}
When the exposure $G$ takes two values zero and one, after a simple calculation
\begin{equation*}
\begin{aligned}
\tau_{canonic}=&\tau+\frac{1}{|D_M|}\sum_{i\in D_M}\Big[\big(\mathbbm{E}(Y_{i2}|W_i=0,G_i=1,z_i)-\mathbbm{E}(Y_{i2}|W_i=0,G_i=0,z_i)\big)\\
&-\big(\mathbbm{E}(Y_{i1}|W_i=0,G_i=1,z_i)-\mathbbm{E}(Y_{i1}|W_i=0,G_i=0,z_i)\big)\Big]\\
&\cdot \big[P(G_i=1|W_i=1,z_i)-P(G_i=1|W_i=0,z_i)\big]
\end{aligned}
\end{equation*}
$\tau_{canonic}$ cannot be interpreted as a total effect. The terms $\mathbbm{E}(Y_{i2}|W_i=0,G_i=1,z_i)-\mathbbm{E}(Y_{i2}|W_i=0,G_i=0,z_i)$ and $\mathbbm{E}(Y_{i1}|W_i=0,G_i=1,z_i)-\mathbbm{E}(Y_{i1}|W_i=0,G_i=0,z_i)$ represent the ``individual spillover effect" defined in Appendix \ref{spilloversection} and the heterogeneity of the first period outcome associated with $G_i$, respectively. Suppose $G_i\not\indep W_i\ |\ z_i$ and there is no first period heterogeneity associated with $G_i$ such that $\mathbbm{E}(Y_{i1}|W_i=0,G_i=1,z_i)-\mathbbm{E}(Y_{i1}|W_i=0,G_i=0,z_i)=0$ $\forall i$, $\tau_{canonic}=\tau$ when there is no ``individual spillover effect" for the non-directly treated units. When the spillover effect is moderate compared with the direct effect, the difference between $\tau_{canonic}$ and $\tau$ can be sizable.

\end{remark}

Following Remark \ref{remark1}, Table \ref{tab:compare} below provides numerical results from a simple simulation study. The detailed data generating process is summarized in Appendix \ref{sec:DGP}. The exposure mapping $G_i$ is correlated with $W_i$ in the design and there is no first period outcome heterogeneity associated with $G_i$. In the first design, when there is no spillover effect on the non-directly treated units, i.e., $\mathbbm{E}(Y_{i2}|W_i=0,G_i=1,z_i)-\mathbbm{E}(Y_{i2}|W_i=0,G_i=0,z_i)=0$, $\forall i$, $\tau_{canonic}$ is almost identical to $\tau$. By contrast, in the second design, there is spillover effect on the non-directly treated units with $\mathbbm{E}(Y_{i2}|W_i=0,G_i=1,z_i)-\mathbbm{E}(Y_{i2}|W_i=0,G_i=0,z_i)=1$, $\forall i$. With the direct effects $\tau(1)=-1$ and $\tau(0)=0$, the spillover effect is comparable to the direct effect in magnitude. In this case, $\tau_{canonic}$ can be substantially different from $\tau$. 
\begin{table}[htbp]  
	\centering  
	\caption{Comparison of $\tau_{canonic}$ and $\tau$} 
	\begin{threeparttable}
		\begin{tabular}{lcc}          
			\toprule
			& \multicolumn{1}{l}{No Spillover} & \multicolumn{1}{l}{With Spillover} \\    
			\midrule
			$\tau_{canonic}$ & 0.764 & -0.223 \\   
			$\tau$   & 0.767 & -0.767 \\    
			\bottomrule
		\end{tabular}  \label{tab:compare}%
		\begin{tablenotes}
			\begin{footnotesize}
				\item[1] In both simulation designs, $G_i\not\indep W_i\ |\ z_i$ and $\mathbbm{E}(Y_{i1}|W_i=0,G_i=1,z_i)-\mathbbm{E}(Y_{i1}|W_i=0,G_i=0,z_i)=0$, $\forall i$.
				\item[2] In the first design, $\tau(1)=1$, $\tau(0)=0$, and  $\mathbbm{E}(Y_{i2}|W_i=0,G_i=1,z_i)-\mathbbm{E}(Y_{i2}|W_i=0,G_i=0,z_i)=0$, $\forall i$; in the second design, $\tau(1)=-1$, $\tau(0)=0$, and $\mathbbm{E}(Y_{i2}|W_i=0,G_i=1,z_i)-\mathbbm{E}(Y_{i2}|W_i=0,G_i=0,z_i)=1$, $\forall i$. 
			\end{footnotesize}
		\end{tablenotes}
	\end{threeparttable}
\end{table}%

\subsection{Modified Two-Way Fixed Effects}
One way to estimate the spillover effect suggested in the existing literature is to augment the TWFE DID estimating equation with another binary indicator $S_{i}$ equal to one if a unit is close to a treated unit; see, for instance, attempts in \cite{di2004police} and \cite{butts2021difference}. Using the notation in the current paper, I modify the estimating equation to be 
\begin{equation}
	\label{twfe}
	Y_{it}=\beta_1 W_{it}+\beta_2 (1-W_i)S_{it}+\beta_3 W_i S_{it}+\alpha_i+\lambda_t+\epsilon_{it},
\end{equation}
where $W_{it}=W_i*\mathbbm{1}\{t=2\}$ and $S_{it}=S_i*\mathbbm{1}\{t=2\}$. $\hat{\beta}_1$ and $\hat{\beta}_1+\hat{\beta}_3-\hat{\beta}_2$ estimated from equation (\ref{twfe}) would be consistent for the DATT defined by 
\begin{equation*}
\bar{\tau}(0)=\frac{1}{|D_M|}\sum_{i\in D_M}\Big[\mathbbm{E}\big(y_{i2}(1,\underline{0})-y_{i2}(0,\underline{0})|W_i=1,S_i=0\big)\Big]
\end{equation*}
and 
\[
\bar{\tau}(1)=\frac{1}{|D_M|}\sum_{i\in D_M}\Big[\mathbbm{E}\big(y_{i2}(1,\bm{W}_{-i})-y_{i2}(0,\bm{W}_{-i})|W_i=1,S_i=1\big)\Big]
\]
respectively, under some parallel trends assumptions. 

We can see that given the estimating equation of the augmented TWFE, the specified exposure mapping is fixed as $\mathbbm{1}\{A_s\bm{W}>0\}=S_i$, where $A_s$ is the spatial weighting matrix with units being neighbors if their distance is less than or equal to $\bar{\rho}$. Only when the interference structure coincides with the indicator function $\mathbbm{1}\{A_s\bm{W}>0\}$ along with the correct distance cutoff, can the augmented TWFE identify the exact direct ATT. In contrast, when the true interference structure is not $\mathbbm{1}\{A_s\bm{W}>0\}$, the proposed estimands in Section \ref{iden_dr} below can still identify the exact direct ATT by choosing correct specification of the exposure mapping with potentially multiple levels of neighborhood exposure $g$.  Meanwhile, covariates can be flexibly accounted for in the proposed estimands by assuming conditional parallel trends.

\begin{remark}
	Inspired by the modified TWFE estimating equation above, for any specification of the exposure mapping one can instead augment TWFE in a saturated way.
	\begin{equation}\label{saturated}
		\begin{aligned}
			Y_{it}=&\beta_0+\beta_1 W_i+\beta_2 W_{it}+\eta_0 (1-W_i)G_{2it}+\eta_1 W_i G_{2it}\\
			&+\delta_0 (1-W_i)G_{3it}+\delta_1 W_i G_{3it}+\cdots+\xi_0 (1-W_i)G_{|\mathcal{G}|it}+\xi_1 W_i G_{|\mathcal{G}|it}+z_i\gamma+\lambda_t+\epsilon_{it},
		\end{aligned}
	\end{equation}
	where one creates $|\mathcal{G}|-1$ binary indicators for each exposure level, $G_{git}=\mathbbm{1}\{G_i=g\}*\mathbbm{1}\{t=2\}$.  
	DATT $\tau(g)$ can be consistently estimated by linear combinations of the coefficient estimators if the linearity in equation (\ref{saturated}) holds true. The saturated TWFE, however, suffers from the same homogeneous (in $z$) restrictions as pointed out by Remark 1 in \cite{sant2020doubly} and lacks flexibility in controlling for covariates. Furthermore, it is possible that some $G_{git}$ may not be well-defined for each unit because $G_i$ cannot take value $g$ for some unit $i$. 
\end{remark}

\subsection{Doubly Robust Estimand}\label{iden_dr}

Since ignoring the spillover effect is only harmless under special scenarios, new estimators need to be proposed for the DATT. Under parallel trends and overlap assumptions, the DATT can be identified by inverse weighting using propensity scores. 
\begin{equation}
\begin{aligned}
\tau^{ipw}(g)=&\frac{1}{|D_M|}\sum_{i\in D_M}\mathbbm{E}\Bigg[\frac{W_i-p(z_i)}{p(z_i)(1-p(z_i))}\frac{\mathbbm{1}\{G_i=g\}}{W_i\pi_{1g}(z_i)+(1-W_i)\pi_{0g}(z_i)}(Y_{i2}-Y_{i1})\bigg|z_i\Bigg]\\
=&\mathbbm{E}_D\Bigg[\frac{W_i-p(z_i)}{p(z_i)(1-p(z_i))}\frac{\mathbbm{1}\{G_i=g\}}{W_i\pi_{1g}(z_i)+(1-W_i)\pi_{0g}(z_i)}(Y_{i2}-Y_{i1})\Bigg]
\end{aligned}
\end{equation}
To simplify notation, I use $\mathbbm{E}_D$ to denote the finite population average conditional on the attributes $\bm{z}$ from now on. 
Without the $G$ indicator and the additional propensity scores for spillover, the IPW-DID estimand is the same as the estimand proposed in \cite{abadie2005semiparametric}. 

Alternatively, the DATT can also be identified through regression adjustment. 
Define the conditional means of the potential outcome as 
\begin{equation}
\label{mean1}
\mu_{t,wg}(z_i)=\mathbbm{E}(Y_{it}|W_i=w,G_i=g,z_i).
\end{equation}
The regression adjustment estimand is 
\begin{equation}
\tau^{reg}(g)=\frac{1}{|D_M|}\sum_{i\in D_M}\Big(\mu_{2,1g}(z_i)-\mu_{2,0g}(z_i)-\big[\mu_{1,1g}(z_i)-\mu_{1,0g}(z_i)\big]\Big).
\end{equation}

To allow for more robustness against misspecification of the propensity scores or the conditional means of the outcomes, the IPW-DID estimand can be extended to an AIPW estimand.
Let $m_{t,wg}(z_i)$ denote the model for equation (\ref{mean1}). Denote $\Delta m_{wg}(z_i)=m_{2,wg}(z_i)-m_{1,wg}(z_i)$. Furthermore, let $\eta(z_i)$, $\eta_{1g}(z_i)$, and $\eta_{0g}(z_i)$ be the models for the propensity scores in equations (\ref{ps})-(\ref{psg0}), respectively. 
The doubly robust estimand is 
\begin{equation}
\label{eq:dr}
\begin{split}
\tau^{dr}(g)=&\mathbbm{E}_D\Bigg[\frac{W_i}{\eta(z_i)}\frac{\mathbbm{1}\{G_i=g\}}{\eta_{1g}(z_i)} \Big(\big(Y_{i2}-m_{2,1g}(z_i)\big)-\big(Y_{i1}-m_{1,1g}(z_i)\big)\Big)\\
&-\frac{1-W_i}{1-\eta(z_i)}\frac{\mathbbm{1}\{G_i=g\}}{\eta_{0g}(z_i)} \Big(\big(Y_{i2}-m_{2,0g}(z_i)\big)-\big(Y_{i1}-m_{1,0g}(z_i)\big)\Big)\\
&+\Delta m_{1g}(z_i)-\Delta m_{0g}(z_i)\Bigg].
\end{split}
\end{equation}
\begin{proposition}\label{pro3}
Under Assumptions \ref{overlap}-\ref{trueparallel}, $\tau^{dr}(g)=\tau(g)$ as long as either $\eta(z)=p(z)$ and $\eta_{wg}(z)=\pi_{wg}(z)$ for $w\in\{0,1\}$ or $\Delta m_{wg}(z)=\mu_{2,wg}(z)-\mu_{1,wg}(z)$ for $w\in\{0,1\}$.  
\end{proposition}
As a result, we can consistently estimate $\tau(g)$ as long as either the models for the propensity scores or the models for the conditional means of the outcome are correctly specified in the doubly robust estimand.

\subsection{Misspecified Exposure Mapping}\label{misspecify}
In this section, I show how to proceed with DID estimation with a chosen $G(\cdot)$ function that is potentially misspecified. For more discussion on some common choices of $G(\cdot)$ and how they can be accommodated in the current framework, I refer readers to Appendix \ref{summary}.   

 In an ideal scenario, empirical researchers would like to come up with a functional form of $G(\cdot)$ that captures actual interactions among neighbors as well as conveys clear causal explanations. Because of the unknown interference structure and the high dimensional treatment assignment vector of the entire population, choosing a $G(\cdot)$ that achieves both goals is challenging.\footnote{\cite{manski2013identification} and \cite{basse2018limitations} formally point out that there exist no consistent treatment effect estimators under arbitrary interference. It is therefore necessary to make dimension reduction assumptions about the interference structure in order to identify meaningful treatment effect parameters.} Recall the example in Section \ref{setup}, $G_i=\sum_{j\in D_M, j\neq i} A_{ij}W_j/\sum_{j\in D_M,j\neq i} A_{ij}$, where $A_{ij}=1$ if the distance between units $i$ and $j$ is within a certain cutoff $\bar{\rho}$. Besides the somewhat arbitrary cutoff $\bar{\rho}$, the impact of $i$'s neighbors may not be exchangeable in reality, e.g., unit $l$ may have greater influence on $i$ compared to unit $m$. That said, the specification above might still capture meaningful policy effects. Consequently, using domain knowledge in the context of each specific empirical question, coming up with a $G(\cdot)$ that summarizes interesting and relevant policy implications for both direct treatment and spillover effects might be a good starting point.

To overcome potential misspecification of the spillover pattern, I consider the expected direct treatment effect at certain neighborhood exposure levels as the parameter of interest inspired by \cite{savje2023causal}. The causal estimands I define coincide with the exact direct treatment effect when the exposure mapping is correctly specified and remain well-defined even under misspecification. 

The parameter of interest is now the expected direct average treatment effect on the treated (EDATT) at exposure levels $g\in \mathcal{G}$, which identifies the direct ATT that would realize in expectation at the specified exposure level.\footnote{One can similarly aggregate EDATT to the overall direct effect.} 
\begin{equation}
\label{eq1}
\begin{aligned}
\tau^*(g)=&\frac{1}{|D_M|}\sum_{i\in D_M}\mathbbm{E}\big[y_{i2}(1,\bm{W}_{-i})-y_{i2}(0,\bm{W}_{-i})|W_i=1, G_i=g, z_i\big]
\end{aligned}
\end{equation}

The key ingredient of the definition is the expected potential outcome at exposure level $g$,
\begin{align*}
&\mathbbm{E}\big[y_{i2}(1,\bm{W}_{-i})|W_i=1, G_i=g, z_i\big]\\
=&\sum_{\bm{w}_{-i}\in\{0,1\}^{|D_M|-1}}\mathbbm{E}\big[y_{i2}(1,\bm{w}_{-i})|W_i=1,\bm{W}_{-i}=\bm{w}_{-i},z_i\big]P(\bm{W}_{-i}=\bm{w}_{-i}|G_i=g,W_i=1,z_i),
\end{align*}
where, in addition to the stochastic potential outcomes, the expectation is taken over all possible realizations of $\bm{W}_{-i}$ given the specified exposure mapping $G(i,\bm{W}_{-i})=g$. The definition of the expected potential outcome is different from what is initially proposed in \cite{savje2023causal}, in which the potential outcome is fixed in an experimental setting and the expectation is with respect to the assignment variables only. Not only that I split the entire treatment vector into $w_i$ and $\bm{w_{-i}}$, but also the stochastic nature of the potential outcomes needs to be taken into account. The randomness of the potential outcome function brings up challenge to causal interpretation of the spillover effect estimand. Appendix \ref{spilloversection} provides more detailed reasoning.   

In order to provide a general framework for identifying EDATT, I impose the following assumption instead. It reduces to Assumption \ref{trueparallel} under correctly specified $G(\cdot)$.\footnote{I consider Assumption \ref{parallel} as an attempt to accommondate misspecification of exposure mapping in practice. Its plausibility depends on the true interference pattern, the chosen exposure mapping, and the design of the individual assignments and hence needs to be examined case by case. If Assumption \ref{parallel} does not hold exactly with misspecified $G(\cdot)$, one might also consider sensitivity analysis along the lines in \cite{rambachan2023more}.} 
\begin{assumptionp}{\ref{trueparallel}$'$}\label{parallel}
(Parallel Trends)
For any $g\in \mathcal{G}$ and $\forall\ i$,
\begin{equation}
\label{pt}
\begin{aligned}
&\mathbbm{E}\big(y_{i2}(0, \bm{W}_{-i})|W_i=1,G_i=g, z_i\big)-\mathbbm{E}\big(y_{i1}(0,\underline{0})|W_i=1,G_i=g, z_i\big)\\
=&\mathbbm{E}\big(y_{i2}(0, \bm{W}_{-i})|W_i=0,G_i=g, z_i\big)-\mathbbm{E}\big(y_{i1}(0,\underline{0})|W_i=0,G_i=g, z_i\big)
\end{aligned}
\end{equation}
\end{assumptionp}

\begin{corollary}\label{coro1}
	Proposition \ref{pro3} holds for $\tau^*(g)$. Namely, $\tau^{dr}(g)=\tau^*(g)$ under Assumptions \ref{overlap}, \ref{anticipation}, \ref{parallel}, and the conditions in Proposition \ref{pro3}. 
\end{corollary}
Corollary \ref{coro1} implies that the same identification results for DATT also hold for EDATT. Consequently, one can use the same estimator proposed in Section \ref{asymptotic} below. Notice that the identification results here and those in Section \ref{iden_dr} hold for any arbitrary $G(\cdot)$, either correctly or incorrectly specified. To show the asymptotic properties of the estimator in a unified framework, I apply the device of approximate neighborhood interference (ANI) in \cite{leung2022causal} to spatial data. The ANI device implies that treatments of units outside of $i$'s $K$-neighborhood can legitimately influence $i$'s potential outcome as long as treatments assigned to units further from $i$ have a smaller, but possibly nonzero, effect on $i$’s response. This device is used to make the asymptotic derivation more tractable but is not required for the identification of the parameter. 

In summary, practitioners can still use the spillover function they choose according to the construction of the exposure mapping $G(\cdot)$ in Section \ref{setup} based on their domain/institutional knowledge. 
Having said that, the actual potential outcome function does not restrict the interference structure exactly as $G(\cdot)$. Based on the same causal effect estimator and a unified asymptotic distribution, practitioners do not need to change their estimation and inference procedure based on their stance on the specification of the spillover pattern. If misspecification of the exposure mapping is a concern, one only needs to modify the interpretation of DATT to EDATT.  

\subsection{Pre-trends}\label{sec:pre}

In empirical research, tests for pre-trends remain common despite the caution described in \cite{roth2022pretest}. A placebo DID is typically applied to multiple periods observed before treatment by imposing a hypothetical period of adoption of treatment. Something similar can be done in the context of interference. Since Assumption \ref{parallel} nests Assumption \ref{trueparallel} when the exposure is correctly specified, I focus on the pre-trends of Assumption \ref{parallel}. In the simplest case, suppose there are two time periods $t=0, 1$ prior to treatment, by imposing the placebo treatment between time periods 0 and 1, we would like to test
\begin{equation}
\label{pre1}
\begin{aligned}
&\frac{1}{|D_M|}\sum_{i\in D_M}\Big[\mathbbm{E}\big(y_{i1}(0, \bm{W}_{-i})|W_i=1,G_i=g, z_i\big)-\mathbbm{E}\big(y_{i0}(0,\underline{0})|W_i=1,G_i=g, z_i\big)\Big]\\
=&\frac{1}{|D_M|}\sum_{i\in D_M}\Big[\mathbbm{E}\big(y_{i1}(0, \bm{W}_{-i})|W_i=0,G_i=g, z_i\big)-\mathbbm{E}\big(y_{i0}(0,\underline{0})|W_i=0,G_i=g, z_i\big)\Big].
\end{aligned}
\end{equation}
The potential outcome $y_{i1}(0, \bm{W}_{-i})$ is not observable because no unit is treated in time period 1. Nevertheless, under the no anticipation assumption, equation (\ref{pre1}) reduces to 
\begin{equation}
\label{pre}
\begin{aligned}
&\frac{1}{|D_M|}\sum_{i\in D_M}\Big[\mathbbm{E}\big(y_{i1}(0, \underline{0})|W_i=1,G_i=g, z_i\big)-\mathbbm{E}\big(y_{i0}(0,\underline{0})|W_i=1,G_i=g, z_i\big)\Big]\\
=&\frac{1}{|D_M|}\sum_{i\in D_M}\Big[\mathbbm{E}\big(y_{i1}(0, \underline{0})|W_i=0,G_i=g, z_i\big)-\mathbbm{E}\big(y_{i0}(0,\underline{0})|W_i=0,G_i=g, z_i\big)\Big].
\end{aligned}
\end{equation}
Namely, one can test whether subgroups defined by the combination of direct treatment status and exposure level have different trends before actual treatment. In reality, the testing equation (\ref{pre}) is a test of both no anticipation and parallel pre-trends.

\section{Asymptotic Properties of the Parametric Estimator}\label{asymptotic}
I am primarily concerned with estimating the DATT (EDATT) in this section. Spillover effects are defined in Appendix \ref{spilloversection}. Their estimation is similar to that of the DATT. I propose a GMM estimator combining equation (\ref{eq:dr}) with moment conditions for the propensity scores and conditional means of outcomes chosen by the empirical researcher. Since the inference is only conditional on the covariates $\bm{z}$, all unobservables are identically distributed conditional on $z_i$.\footnote{It suffices that the first conditional moment is identical.} The inference framework implies that the individual propensity score function and the individual conditional mean function of the outcome remain the same across units. 

To make the estimators more robust to misspecification of these functions, one can use various moment conditions to identify the propensity scores. One option is the covariate balancing propensity scores (CBPS) in \cite{imai2014covariate} or similarly the inverse probability tilting estimator in \cite{graham2012inverse}, which can be locally more robust than the propensity scores based on maximum likelihood estimation (MLE).\footnote{The alternative would be estimating all functions semiparametrically or nonparametrically, which is left as future work.}

I denote a generic moment condition for propensity scores as 
\begin{equation}
\mathbbm{E}_D \big[q_1(W_i,z_i,\gamma^*_1)\big]=\bf{0}
\end{equation}
and 
\begin{equation}
\mathbbm{E}_D \big[q_2(W_i,G_i, z_i,\gamma^*_2)\big]=\bf{0},
\end{equation}
where $z_i$ can contain neighbors' attributes. 
For instance, the moment conditions for CBPS are
\begin{equation}\label{eq:m1}
\mathbbm{E}_D\bigg[\frac{W_i}{P(W_i=1|z_i)}z_i-\frac{(1-W_i)}{1-P(W_i=1|z_i)}z_i\bigg]=\bf{0}
\end{equation}
and for $g=1,2,\dots,G-1$,
\begin{equation}\label{eq:m2}
\mathbbm{E}_D\bigg[\frac{\mathbbm{1}\{G_i=g\}}{P(G_i=g|W_i,z_i)}(W_i,z_i)-\frac{\mathbbm{1}\{G_i=g-1\}}{P(G_i=g-1|W_i,z_i)}(W_i,z_i)\bigg]=\bf{0},
\end{equation}
where $P(W_i=1|z_i)$ is some probability for a binary response, such as $\frac{exp(z_i\gamma^*_1)}{1+exp(z_i\gamma^*_1)}$, and $P(G_i=g|W_i,z_i)$ is some probability for discrete choices. 
Similarly, generic conditional moment conditions are denoted by 
\begin{equation}\label{eq:m3}
\mathbbm{E}_D \big[q_3(Y_{i1},W_i,G_i,z_i,\gamma^*_3)\big]=\bf{0}
\end{equation}
and 
\begin{equation}\label{eq:m4}
\mathbbm{E}_D \big[q_4(Y_{i2},W_i,G_i,z_i,\gamma^*_4)\big]=\bf{0}.
\end{equation}
Alternatively, one can model the conditional mean for $\Delta Y_i=Y_{i2}-Y_{i1}$ and formulate the moment condition as 
\begin{equation}
\mathbbm{E}_D \big[\tilde{q}_{3}(\Delta Y_i,W_i,G_i, z_i,\tilde{\gamma}^*_3)\big]=\bf{0}.
\end{equation}
Leading cases for outcome regression are moment conditions from (nonlinear) least squares.
Lastly, the moment condition for $\tau(g)$ is a restatement of equation (\ref{eq:dr})\footnote{In practice, it is recommended to normalize the weights for IPW type estimators. Changing the moment condition with normalized propensity scores -- where the weights sum to unity -- does not affect asymptotic normality of the GMM estimator. In fact, estimators with normalized weights consistently show better finite sample performance in the simulations below.}. Denote $\theta^*_M=({\gamma^*_1}',{\gamma^*_2}',{\gamma^*_3}',{\gamma^*_4}',\tau(g))'$.
\begin{align*}
&\mathbbm{E}_D\big[q_5(Y_{it},W_i,G_i,z_i,\theta^*_M)\big]\\
=&\mathbbm{E}_D\Bigg[\frac{W_i}{\eta(z_i)}\frac{\mathbbm{1}\{G_i=g\}}{\eta_{1g}(z_i)} \Big(\big(Y_{i2}-m_{2,1g}(z_i)\big)-\big(Y_{i1}-m_{1,1g}(z_i)\big)\Big)\\
&-\frac{1-W_i}{1-\eta(z_i)}\frac{\mathbbm{1}\{G_i=g\}}{\eta_{0g}(z_i)} \Big(\big(Y_{i2}-m_{2,0g}(z_i)\big)-\big(Y_{i1}-m_{1,0g}(z_i)\big)\Big)\\
&+\Delta m_{1g}(z_i)-\Delta m_{0g}(z_i)-\tau(g)\Bigg]=0
\tag{\stepcounter{equation}\theequation}
\end{align*}

Let $X_i=\{Y_{it},W_i,G_i, z_i\}$, $q(X_i,\theta)=(q_1'(\gamma_1),q_2'(\gamma_2),q_3'(\gamma_3),q_4'(\gamma_4),q_5(\theta))'$, and $\widehat{\Psi}$ be the weighting matrix with dimensions larger or equal to that of $\theta$.
\begin{equation}\label{gmm}
\hat{\theta}=\arg\min_{\theta\in \Theta}\frac{1}{|D_M|}\sum_{i\in D_M}q(X_i,\theta)'\widehat{\Psi}\frac{1}{|D_M|}\sum_{i\in D_M}q(X_i,\theta)
\end{equation}
The GMM estimator is the solution to the finite population minimization problem in equation (\ref{gmm}). And the estimator of $\tau(g)$ is the last element of $\hat{\theta}$. Notice that the same estimation procedure applies to $\tau^*(g)$.
I impose the following assumptions to study the asymptotic distribution of the GMM estimator.
\begin{assumption}\label{ap:domain}
Suppose $\{D_M\}\subseteq \mathbb{R}^d$, $d\geq 1$ is a sequence of finite sets such that $|D_M|\to \infty$ as $M\to \infty$. All elements in $D_M$ are located at distances of at least $\rho_0>0$ from each other, i.e., for all $i,j\in D_M$: $\rho(i,j)\geq \rho_0$; w.l.o.g. I assume that $\rho_0>1$. 
\end{assumption}

I adopt the increasing domain asymptotics with spatial data as implied by Assumption \ref{ap:domain}. The assumption of the minimum distance ensures the expansion of the finite population region in our asymptotic framework when the population size keeps growing. This rules out the case where population size grows within a bounded population region in the sense that units become arbitrarily dense in a given region.

\begin{assumption}\label{ap:ani2}(Approximate Neighborhood Interference)
Let $\bm{W}^{(i,s)}=\big(\bm{W}_{\mathcal{N}(i,s)}, \bm{W}'_{D_M\backslash \mathcal{N}(i,s)}\big)$, where $\bm{W}'$ is an independent copy of $\bm{W}$, $\bm{W}^{(i,s,0)}=\big(\bm{W}_{\mathcal{N}(i,s)}, \underline{0}\big)$, i.e., $\bm{W}'_{D_M\backslash \mathcal{N}(i,s)}=\underline{0}$, and
\begin{equation}\label{anidef}
\kappa_M(s)=\max_{i\in D_M}\mathbbm{E}\Big[\big|y_{i2}(\bm{W})-y_{i2}\big(\bm{W}^{(i,s,0)}\big)\big|\Big|\bm{z}\Big].
\end{equation}
Suppose that $\sup_{M}\kappa_M(s)\to 0$ as $s\to \infty$.
\end{assumption}

Assumption \ref{ap:ani2} is a modified version of Assumption 4 in \cite{leung2022causal}. \cite{leung2022causal} varies $\bm{W}'_{D_M\backslash \mathcal{N}(i,s)}$ in an arbitrary way but these treatments outside of the $s$-neighborhood are fixed at zero here. Assumption \ref{ap:ani2} essentially implies that treatments of units from $s$ distance away from $i$ should have a minimal impact as the distance $s$ gets larger. This way, we can allow interference from outside the immediate $K$-neighborhood while still being able to derive the asymptotic properties of the proposed estimators. With that said, interference is restricted in a way that a unit's exposure is primarily, but not entirely, determined by the assignments of neighbors closer to it.

Needless to say, Assumption \ref{ap:ani2} is automatically satisfied under correctly specified exposure mapping within the $K$-neighborhood. We can also accommodate misspecified exposure mapping that satisfies Assumption \ref{ap:ani2} in our asymptotic framework. 
Appendix \ref{summary} gives an overview of different approaches to modeling interference taken in the literature and compares them to ANI.

I adopt $\psi$-dependence in \cite{kojevnikov2021limit} as the notion of weak dependence throughout the paper. Notice that $\alpha$-mixing is a special case of $\psi$-dependence. 
Let $\mathcal{L}_{\nu, h}$ denote the collection of bounded Lipschitz real functions $f(\cdot)$ on $\mathbb{R}^{\nu\times h}$ with the Lipschitz constant $\text{Lip}(f)<\infty$ and $\norm{f}_\infty<\infty$, where $\norm{f}_\infty=\sup_x|f(x)|$. 
Denote the collection of subset pairs as 
\[
\mathcal{P}_M(h,h';s)=\{(H,H'):H,H'\subseteq D_M, |H|=h,|H'|=h', \rho(H,H')\geq s\}.
\]
\begin{definition}\label{def}
	A triangular array $\{V_i, i\in D_M,M\geq 1\}, V_i\in \mathbb{R}^\nu$, is called $\psi$-dependent if there exist uniformly bounded constants $\{\tilde{\kappa}_{M,s}\}_{s\geq 0}$ with $\tilde{\kappa}_{M,0}=1$, and a collection of nonrandom functions $\{\psi_{h,h'}\}_{h,h'\in \mathbb{N}}$ with $\psi_{h,h'}:\mathcal{L}_{\nu,h}\times \mathcal{L}_{\nu,h'}\to [0,\infty)$ such that for all $(H,H')\in \mathcal{P}_M(h,h';s)$ with $s>0$ and all $f\in \mathcal{L}_{\nu,h}$ and $f'\in \mathcal{L}_{\nu,h'}$,
	\begin{equation*}\label{eq:def}
		\big|Cov\big(f(V_H),f'(V_{H'})|\bm{z} \big)\big|\leq \psi_{h,h'}(f,f')\tilde{\kappa}_{M,s},
	\end{equation*}
	where $V_H=(V_i:i\in H)$.
\end{definition}

I require $\tilde{\kappa}_{M,s}$ to approach zero as $s$ grows. $\psi$-dependence bounds the covariances of any two subsets of observations distant from each other. 

\begin{assumption}\label{ap:mixing}
Let $Y_{it}=h(W_i,\bm{W}_{-i},z_i,U_{it})$, where $h(\cdot)$ is some generic function and $U_{it}$ denotes the unobservables. Let $\epsilon_i=(W_i,U_{i1},U_{i2})$. The random field $\epsilon=\{\epsilon_i,i\in D_M, M\geq 1\}$ is $\alpha$-mixing under Definition 2 in \cite{jenish2012spatial}. The mixing coefficient is denoted by $\alpha^\epsilon(u,v,r)\leq (u+v)\widehat{\alpha}^\epsilon(r)$.
\end{assumption}

On top of possible interference, Assumption \ref{ap:mixing} allows assignment variables to be spatially correlated as well. 

\begin{lemma}\label{weakdependence}
Under Assumptions \ref{ap:domain}, \ref{ap:ani2}, \ref{ap:mixing}, and Assumption \ref{ap:gmm} in Appendix A, for each $\theta\in\Theta$, each element of $q(X_i,\theta)$ and $\nabla_\theta q(X_i,\theta)$ is $\psi$-dependent with $\tilde{\kappa}_{M,s}=\big(\kappa_M(s/3)+s^d\widehat{\alpha}^{\epsilon}(s/3)\big)\mathbbm{1}(s> 3\max\{K,\rho_0\})+\mathbbm{1}(s\leq 3\max\{K,\rho_0\})$. 
\end{lemma}

Lemma \ref{weakdependence} shows that Assumption \ref{ap:ani2} together with Assumption \ref{ap:mixing} ensure that the data is weakly dependent even with spatially correlated assignment and spillover effect, regardless of the correct specification of the exposure mapping. 

To adapt the limit theorems in \cite{kojevnikov2021limit} to spatial data, I replace the network denseness with the cardinality of the spatial sets implied by Lemma A.1 in \cite{jenish2009central}. 
As a result, 
Assumption 3.2 in \cite{kojevnikov2021limit} is modified as 
\begin{assumption}\label{ap:wlln}
\[
\sum^{\infty}_{s=1}s^{d-1}\tilde{\kappa}_{M,s}<\infty
\]
\end{assumption}
Assumption \ref{ap:wlln} is in line with Assumption 3(b) in \cite{jenish2009central} for $\alpha$-mixing random fields. It imposes a trade-off between the sizes of $s$-neighborhood boundaries and the rate of decay of spatial dependence.  
Let $\sigma_M^2=Var\big[\sum_{i\in D_M}\lambda'q(X_i,\theta_M^*)|\bm{z}\big]$ for a nonzero vector $\lambda$.
Similarly, Assumption 3.4 in \cite{kojevnikov2021limit} is modified as 
\begin{assumption}\label{ap:clt}
There exists a positive sequence $r_M\to\infty$ such that for $k=1,2$
\[
\frac{1}{\sigma_M^{2+k}}\sum_{i\in D_M}\sum^{\infty}_{s=0}s^{d-1}\max_{j\in D_M, s\leq \rho(i,j)<s+1}\big|\mathcal{N}(i;r_M)\setminus \mathcal{N}(j;s-1)\big|^k\tilde{\kappa}_{M,s}^{1-\frac{2+k}{p}}\to 0
\]
and
\[
\frac{|D_M|^2\tilde{\kappa}_{M,r_M}^{1-(1/p)}}{\sigma_M}\to 0
\]
as $M\to \infty$, where $p>4$ is that appears in Assumption \ref{ap:gmm} in Appendix \ref{sec:proof}.
\end{assumption}
The rate of $\tilde{\kappa}_{M,s}$ is implicitly implied by Assumption \ref{ap:clt}, which restricts the spatial sets and the cross-sectional dependence of random variables. A sufficient condition for the first part of the assumption is
\[
\frac{|D_M|}{\sigma_M^{2+k}}r_M^{kd}\sum^{\infty}_{s=0}s^{d-1}\tilde{\kappa}_{M,s}^{1-\frac{2+k}{p}}\to 0.
\]
Analogous conditions can be found in \cite{jenish2009central} as equations (B.18) and (B.19) therein.

The notation used in the asymptotic distribution of the GMM estimator is introduced as follows. 
Define
\begin{equation}\label{eq:variance}
\Omega_M=\Delta_{ehw,M}+\Delta_{spatial,M}-\Delta_{E,M}-\Delta_{ES,M},
\end{equation}
where 
\begin{equation}
\Delta_{ehw,M}=\frac{1}{|D_M|}\sum_{i\in D_M}\mathbb{E}\big[q(X_i,\theta^*_M)q(X_i,\theta^*_M)'|\bm{z}\big],
\end{equation}
\begin{equation}
\Delta_{E,M}=\frac{1}{|D_M|}\sum_{i\in D_M}\mathbb{E}\big[q(X_i,\theta^*_M)|\bm{z}\big]\mathbb{E}\big[q(X_i,\theta^*_M)|\bm{z}\big]',
\end{equation}

\begin{equation}
\Delta_{spatial,M}=\frac{1}{|D_M|}\sum_{i\in D_M}\sum_{j\in D_M, j\neq i}\mathbb{E}\big[q(X_i,\theta^*_M)q(X_j,\theta^*_M)'|\bm{z}\big],
\end{equation}
and
\begin{equation}
\Delta_{ES,M}=\frac{1}{|D_M|}\sum_{i\in D_M}\sum_{j\in D_M, j\neq i}\mathbb{E}\big[q(X_i,\theta^*_M)|\bm{z}\big]\mathbbm{E}\big[q(X_j,\theta^*_M)|\bm{z}\big]'.
\end{equation}
$\Delta_{ehw,M}$ and $\Delta_{spatial,M}$ account for heteroskedasticity and spatial correlation respectively, whereas $\Delta_{E,M}$ and $\Delta_{ES,M}$ are their finite population counterparts. 
Denote
\[ 
R_M^*=\mathbbm{E}_D\big[\nabla_\theta q(X_i,\theta^*_M)\big]
\]
and
\begin{equation}
V_M=\big({R_M^*}'\Psi_M R_M^*\big)^{-1}{R_M^*}'\Psi_M \Omega_M \Psi_M R_M^* \big({R_M^*}'\Psi_M R_M^*\big)^{-1},
\end{equation}
where $\widehat{\Psi}-\Psi_M\overset{p}\to \textbf{0}$.
\begin{theorem}\label{thm:clt}
Under Assumptions \ref{overlap}, \ref{anticipation}, \ref{parallel}, \ref{ap:domain}-\ref{ap:clt}, and Assumption \ref{ap:gmm} in Appendix \ref{sec:proof}, if either equations (\ref{ps})-(\ref{psg0}) or $\mu_{2,wg}(z)-\mu_{1,wg}(z)$ in equation (\ref{mean1}) are correctly specified, 
\[
V_M^{-1/2}\sqrt{|D_M|}(\hat{\theta}-\theta^*_M)\overset{d}\to \mathcal{N}(\textbf{0},I_k).
\]
\end{theorem}

With the consideration of interference and potentially spatially correlated assignments, we need to make inference robust to spatial correlation.
As a common approach to adjust the variance estimator for spatial correlation, the usual spatial heteroskedasticity and autocorrelation consistent (SHAC) variance estimator is defined as 
\[
\hat{V}=\big(\hat{R}'\widehat{\Psi}\hat{R}\big)^{-1}\hat{R}'\widehat{\Psi}\tilde{\Omega}(\hat{\theta})\widehat{\Psi}\hat{R} \big(\hat{R}'\widehat{\Psi}\hat{R}\big)^{-1},
\]
where 
\[
\hat{R}=\frac{1}{|D_M|}\sum_{i\in D_M}\nabla_\theta q(X_i,\hat{\theta})
\]
and 
\[
\tilde{\Omega}(\theta)=\frac{1}{|D_M|}\sum_{s=0}^{\infty}\omega\Big(\frac{s}{b_M}\Big)\sum_{i\in D_M}\sum_{j\in D_M, s\leq \rho(i,j)< s+1}q(X_i,\theta)q(X_j,\theta)'
\]
with $b_M$ being the bandwidth parameter.
I impose the following assumption for the estimation of the variance-covariance matrix. 
\begin{assumption}\label{ap:variance}
The weights satisfy: \\
(\romannumeral 1) $\omega (0)=1$, $\omega\big(\frac{s}{b_M}\big)=0$ for any $s>b_M$, $\big|\omega\big(\frac{s}{b_M}\big)\big|<\infty$, $\forall$ $M$; \\
(\romannumeral 2)
\[\sum^{\infty}_{s=1}\Big|\omega\Big(\frac{s}{b_M}\Big)-1\Big|s^{d-1}\tilde{\kappa}_{M,s}^{1-2/p}\to 0;\]
(\romannumeral 3) 
\[
\frac{1}{|D_M|}\sum^{\infty}_{s=0}s^{d-1}b_M^{2d}\tilde{\kappa}_{M,s}^{1-4/p}\to 0
\]
as $M\to \infty$, where $b_M=o\big(|D_M|^{1/2d}\big)$ and $p>4$ is that appears in Assumption \ref{ap:gmm} in Appendix \ref{sec:proof}. 
\end{assumption}

Assumption \ref{ap:variance}(\romannumeral 1) is satisfied by common choices of kernels, including the Bartlett and Parzen kernels. Assumption \ref{ap:variance}(\romannumeral 2) requires that the kernel weights $\omega\big(\frac{s}{b_M}\big)$ converge to one sufficiently fast as $M\to\infty$. Part (\romannumeral 3) of Assumption \ref{ap:variance} regulates the growth rate of the bandwidth $\{b_M\}$.

\begin{theorem}\label{thm:variance}
Under conditions in Theorem \ref{thm:clt} and Assumption \ref{ap:variance}, 
\[
\hat{V}-(V_M+V_E)\overset{p}\to \textbf{0},
\] 
where 
\[
V_E=\big({R_M^*}'\Psi_M R_M^*\big)^{-1}{R_M^*}'\Psi_M \Omega_E \Psi_M R_M^* \big({R_M^*}'\Psi_M R_M^*\big)^{-1}\]
and 
\[
\Omega_E=\frac{1}{|D_M|}\sum_{s=0}^{\infty}\omega\Big(\frac{s}{b_M}\Big)\sum_{i\in D_M}\sum_{j\in D_M, s\leq \rho(i,j)< s+1}\mathbb{E}\big[q(X_i,\theta^*_M)|\bm{z}\big]\mathbb{E}\big[q(X_j,\theta^*_M)|\bm{z}\big]'.
\]
\end{theorem}

\begin{remark}
	I state Theorems \ref{thm:clt} and \ref{thm:variance} in terms of Assumption \ref{parallel} so that the asymptotic properties hold for the estimator for both $\tau(g)$ and $\tau^*(g)$. Thus, the same estimation and inference procedure applies to the DATT or EDATT. Only the interpretation of the direct treatment effect needs to be modified if the exposure mapping is suspected to be misspecified. 
	\end{remark}

\begin{remark}\label{remark2}
	When we choose kernel functions that produce positive semi-definite (PSD) weighting matrix, the usual SHAC variance estimator is generally conservative for the finite population conditional spatial-correlation robust variance-covariance matrix.
\end{remark}

The conservativeness of the usual variance estimator for conditional variance has also been investigated in \cite{abadie2014inference} under the independence assumption for the heteroskedasticity-robust variance matrix. I extend it to the case with spatial correlation here when $\Omega_E$ is PSD based on a PSD kernel weighting matrix. An exception to Remark \ref{remark2} is when $\mathbbm{E}\big[q(X_i,\theta^*_M)|\bm{z}\big]=\bf{0}$ for all $i\in D_M$. In this case, the usual variance-covariance matrix estimator is no longer conservative as $V_E=\bf{0}$. With heterogeneous direct treatment effect or misspecification of either the propensity scores or conditional means, $\mathbbm{E}\big[q(X_i,\theta^*_M)|\bm{z}\big]\neq\bf{0}$. 

That said, I would like to highlight a few points. First, because $\tilde{\Omega}(\hat{\theta})$ is a conservative estimator for $\Omega_M$, even if we choose $\Psi_M$ as the optimal weighting matrix $\Omega_M^{-1}$, using $\widehat{\Psi}=\tilde{\Omega}(\hat{\theta})^{-1}$ in estimation is not going to achieve the most efficient GMM estimator. The usual variance estimator is therefore conservative not only because of the neglect of the additional terms in the variance-covariance matrix but also because the optimal weighting matrix is not consistently estimated. Of course, when the model is just identified, the weighting matrix choice is irrelevant. 

Second, unlike the finite population variance-covariance matrix in \cite{xu2022design}, the conditional spatial-correlation robust variance matrix is consistently estimable because it is no longer conditional on the unobserved potential outcomes. There are different approaches one can take. However, since the usual SHAC variance estimator is known to suffer from downward bias especially when the spatial correlation is high, it is not always necessary to estimate the smaller conditional variance matrix.  

\section{Simulations}\label{sec:simulation}
In the simulation, I show the finite sample performance of the proposed estimators for DATT (EDATT). I consider an irregularly spaced lattice with $|D_M|=400$ units. The locations ($s_{1,iM}$,$s_{2,iM}$) are drawn once and kept fixed across replications. Each of $s_{1,iM}$ and $s_{2,iM}$ is independently drawn from $\mathcal{U}(0,20)$. The distance between units $i$ and $j$ is measured by $\rho(i,j)=\max\{|s_{1,iM}-s_{1,jM}|,|s_{2,iM}-s_{2,jM}|\}$. Units are considered neighbors if $\rho(i,j)\leq 0.3$ with the neighborhood structure summarized by the normalized contiguity matrix, $A$. After ruling out units without neighbors, the effective size of the subpopulation eligible for spillover reduces to 350. 

I consider two time period panel data. The potential outcome function in the first time period remains the same across different designs. 
\[
y_1(0,\underline{0})=1+z+e_1,
\]
where $z$ is the individual covariate independently drawn from the standard normal distribution and kept fixed, while $e_1$ is the first time period unobservable. There is a single binary treatment variable $W=\mathbbm{1}\{p(z^*)>u\}$ with $u_i\overset{i.i.d}\sim \mathcal{U}(0,1)$. I vary the second time period potential outcome function and the assignment probability $p(z^*)$ in different designs summarized in Table \ref{tab:design} below. $z^*=(z,z_u)$, where the vector of $z_u$ in the assignment probability is drawn from a multivariate normal distribution with mean zero and a variance-covariance matrix equal to 0.5 raised to the power of the distance between units. Thus, $z_u$ is a spatially correlated locational covariate that stands for neighborhood similarity, which might be neglected in naive estimation assuming away spillover effect. Along with the individual second time period unobservable $e_2$, $e_{i1}|W_i,W_{-i},z_i\sim \mathcal{N}(W_i*z_i,1)$, $e_{i2}|W_i,W_{-i},z_i\sim \mathcal{N}(W_i*z_i,1)$ and $e_{i1}\indep e_{i2}|W_i,W_{-i},z_i$, $\forall\ i$. The specified exposure mapping is denoted by $G=\mathbbm{1}\{A\bm{W}>0\}$, which may or may not coincide with the true interference structure.

\begin{table}[htbp]\label{sdesign}
	\centering
	\caption{Simulation Designs} 
	\begin{threeparttable}
		\begin{tabular}{ccl} 
			\toprule
			Design    & Assignment probability & Second period potential outcomes\\
			\midrule
			1     & $p_1$ & $Y_2=2+W+G+z+e_2$ \\
			2      &  $p_1$   &  $Y_2=2+W+G+2z+e_2$ \\
			3       & $p_2$  & $Y_2=2+W+G+2z+e_2$ \\
			4       & $p_2$  & $\bm{Y}_2=2+\bm{W}+0.2A*\bm{Y}_2+2\bm{z}+\bm{e}_2$ \\
			5 & $p_2$  &  $\bm{Y}_2=2+\bm{W}+0.2A*\bm{Y}_2+2\bm{z}^2+\bm{e}_2$ \\
			6 & $p_2$ &  $Y_2=2+W*G+2z+e_2$\\
			\bottomrule
		\end{tabular}
		\begin{tablenotes}
			\begin{footnotesize}
				\item[1] $p_1=p(z)=\frac{exp(0.3z)}{1+exp(0.3z)}$; $p_2=p(z^*)=\frac{exp(0.3z+0.8z_u)}{1+exp(0.3z+0.8z_u)}$.
				\item[2] $\bm{Y}_2$, $\bm{W}$, $\bm{z}$, and $\bm{e}_2$ stand for the $M\times 1$ vector of $Y_2$, $W$, $z$, and $e_2$.
				\item[3] In designs 4 and 5, the parallel trends assumption holds approximately with the difference between the trends being less than 0.001.
			\end{footnotesize}
		\end{tablenotes}
	\end{threeparttable}
	\label{tab:design}
	
\end{table}

I compare the following estimators: the canonical TWFE, Abadie's IPW estimator, the augmented TWFE, regression adjustment, IPW estimator with either MLE or CBPS moment condition for the propensity scores, and the proposed AIPW estimator with either MLE or CBPS moment condition for the propensity scores. Appendix \ref{sec:addsim} contains the standard deviation of the proposed estimators and the coverage rate of the 95\% confidence intervals based on the usual SHAC standard errors for the doubly robust estimators.

For the canonical Abadie's IPW estimator, I only include $z$ in the logit model of $W$ as interference is assumed away when employing the canonical DID estimators. As an illustration of Proposition \ref{pro1}, I also report Abadie's IPW estimator with $z$, $A\bm{z}$, and $z_u$ included in the logit model, which leads to conditional independence of $W$ and $G$. The estimation of the augmented TWFE follows equation (\ref{twfe}) with $S_i=G_i$. For the estimation of the proposed IPW, regression, and AIPW estimator accounting for spillover effect, the propensity scores for $W$ and $G$ are estimated based on a logit model on $z$, $A\bm{z}$, and $z_u$ and a logit model on $W$, $z$, $A\bm{z}$, and $z_u$, respectively. For the first time period data, I regress $Y_1$ on $W$, $z$, and $W*z$. As for the second period data, I regress $Y_2$ on $W$, $z$, $W*z$, and $G$. All estimators involving weighting are weighted by the normalized propensity scores. The results are summarized across 10,000 replications.

\begin{table}[htbp]
	\centering
	\caption{Direct ATT}
	\begin{threeparttable}
		\begin{tabular}{lcccccc}
			\toprule
			& 1     & 2     & 3     & 4     & 5     & 6 \\ 
			\midrule
			twfe  & 0.998 & 1.259 & 1.355 & 1.333 & 0.884 & 0.909 \\
			abadie($z$) & 0.999 & 1.000 & 1.098 & 1.069 & 1.147 & 0.654 \\
			abadie($z^*$) & 0.997 & 0.997 & 0.999 & 1.028 & 1.031 & 0.607 \\
			atwfe1 & 1.003 & 1.272 & 1.250 & 1.289 & 0.842 & 1.250 \\
			atwfe0 & 0.998 & 1.248 & 1.270 & 1.296 & 0.771 & 0.270 \\
			ra1   & 1.001 & 1.001 & 1.001 & 1.041 & 1.139 & 0.692 \\
			ra0   & 1.001 & 1.001 & 1.001 & 1.041 & 1.139 & 0.692 \\
			ipw\_mle1  & 0.999 & 0.998 & 1.001 & 1.028 & 1.034 & 1.001 \\
			ipw\_mle0  & 1.007 & 1.018 & 1.042 & 1.088 & 1.083 & 0.042 \\
			ipw\_cbps1 & 0.998 & 0.996 & 1.000 & 1.026 & 1.038 & 1.000 \\
			ipw\_cbps0 & 1.009 & 1.019 & 1.037 & 1.081 & 1.076 & 0.037 \\
			dr\_mle1 & 1.002 & 1.002 & 1.002 & 1.028 & 1.037 & 1.001 \\
			dr\_mle0 & 0.997 & 0.997 & 0.999 & 1.045 & 1.084 & 0.001 \\
			dr\_cbps1 & 1.002 & 1.002 & 1.002 & 1.028 & 1.040 & 1.001 \\
			dr\_cbps0 & 0.997 & 0.997 & 0.999 & 1.042 & 1.072 & 0.001 \\
			\bottomrule
		\end{tabular}%
		\begin{tablenotes}
			\begin{footnotesize}
				\item[1] twfe stands for the canonical TWFE estimator; abadie($z$) stands for the canonical Abadie's IPW estimator including only $z$ as the covariate; abadie($z^*$) stands for the canonical Abadie's IPW estimator using $z$, $A\bm{z}$, and $z_u$ as the covariates; atwfe stands for the augmented TWFE estimator; ra stands for the regression adjustment estimator; ipw\_mle stands for the proposed IPW estimator with MLE moment condition for the propensity scores; ipw\_cbps stands for the proposed IPW estimator with CBPS moment condition for the propensity scores; dr\_mle stands for the proposed doubly robust estimator with MLE moment condition for the propensity scores; dr\_cbps stands for the proposed doubly robust estimator with CBPS moment condition for the propensity scores; 
				\item [2] All estimators ending with 1 or 0 correspond to the estimator for the direct treatment effect at exposure levels one or zero, respectively.     
				\item [3] $\tau(1)=\tau(0)=1$ in designs 1-5; $\tau(1)=1$, $\tau(0)=0$ in design 6 with the overall direct effect being approximately 0.607.
			\end{footnotesize}
		\end{tablenotes}
	\end{threeparttable}
	\label{tab:2}%
\end{table}%

According to the population generating process, the direct effects are $\tau(1)=\tau(0)=1$ in designs 1-5 and $\tau(1)=1$, $\tau(0)=0$ in design 6. In the last design, the overall direct effect is approximately 0.607. The point estimates for the direct effect are summarized in Table \ref{tab:2}. In designs 1 and 2, neighborhood similarity does not drive treatment assignment. As a result, the canonical Abadie's IPW estimator with covariate $z$ closely estimates the overall direct effect. The canonical TWFE only performs well in design 1 as the estimating equation of TWFE rules out $z$-specific time trends, which is violated in all other designs.  
The augmented TWFE estimators suffer from the same linearity restriction in their estimating equation as the regular TWFE. With the inclusion of both $z$ and $z_u$, Abadie's IPW estimates are very close to the overall direct effect. 

The proposed estimators accounting for the spillover effects all perform relatively well. Due to the specific exposure mapping functional form, the overlap condition holds better for exposure level one than zero. Consequently, the point estimates for the direct effect estimator at exposure level one are slightly more accurate than the results for the estimator at exposure level zero. It is worth mentioning that the propensity score model for $G$ is always misspecified. The outcome regressions are also misspecified in designs 4-6. Nevertheless, the estimates from the proposed IPW and doubly robust estimators are all quite close to the truth and much more accurate than the TWFE type estimators. 

The doubly robust estimators improve upon regression adjustment and IPW alone, especially at exposure level zero. The only exception is design 5. Since the outcome regression is more severely misspecified than in other designs, we do not see improvement moving from IPW to AIPW. Nevertheless, the AIPW estimates are still better than the regression adjustment estimates. Estimators with CBPS moment condition slightly improve upon estimators with MLE moment condition. When the overlap condition holds weaker in other population generating processes, for instance, changing the assignment probability to $p(z^*)=\frac{exp(z+2z_u)}{1+exp(z+2z_u)}$, we can see more noticeable improvement from using the CBPS moment condition instead of the MLE moment condition. Moreover, the doubly robust estimator can perform substantially better than the proposed IPW estimator at exposure level zero.

\section{Empirical Illustration}\label{sec:application}
I evaluate the effects of China's special economic zones (SEZ) policy using the proposed doubly robust estimators. SEZs are a prominent development strategy that aims to foster agglomeration economies. The benefits of SEZs include corporate tax concessions, customs duty exemptions, discounts on land use fees, and special bank loan programs. SEZs are likely to affect neighboring non-SEZ areas through, for instance, firm relocation or knowledge spillover. 

The data for the empirical illustration come from \cite{lu2019place}. There are five waves of SEZ establishment in China. Each wave is different in nature and targets different regions with earlier waves creating more national-level economic zones.\footnote{As a result, SEZ establishment in China cannot simply be considered as staggered adoption.} Since detailed village level data is only available starting from 2004, \cite{lu2019place} focus on the latest wave of SEZs established between 2005 and 2008. China established 663 SEZs at the provincial level in 2006, accounting for 42 percent of the country's SEZs. These SEZs cover the coastal, central, and western regions and are considered small-scale regional SEZs. As a result, the policy effect is interpreted as the treatment effect on villages that had not yet been treated prior to this wave. This means that areas covered by zones from earlier waves are not included in the finite population. 

\cite{lu2019place} collect comprehensive data on China's economic zones based on the economic censuses conducted by China's National Bureau of Statistics in 2004 and 2008 covering all manufacturing firms. Consequently, the entire finite population of village-level data in 2004 and 2008 is observed, where 2004 is the period prior to the treatment and 2008 post the treatment. The units of observation are villages, which are the most disaggregated geographical units and smaller than an SEZ. Treated villages are referred to as SEZ villages. Unfortunately, the publicly available data from \cite{lu2019place} do not contain an identifier of villages nor distances among villages. Nevertheless, I can match counties in which villages are located from separate datasets published by \cite{lu2019place}. In the organized dataset, there are 3,963 SEZ villages and 99,259 non-SEZ villages. It would be ideal to set certain neighborhood boundaries for each village based on the distance between villages. The exposure mapping is then a function of neighboring villages' SEZ assignment status. In the absence of detailed geographical data, I consider each village's neighborhood to be its corresponding county, with its neighbors being the other villages within the same county. Without distance measures, standard errors are clustered at the county level, which can be considered a special case of spatial-correlation robust inference.  

According to equation (\ref{eq:dr}), the outcome variables $Y_{it}$ include the logarithm of capital, employment, and output of firms in a village. The direct treatment variable $W_i$ is equal to one if village $i$ is located within the boundaries of SEZs and zero otherwise. As an illustration, exposure mapping is defined in the following way. I define SEZ ratio as the fraction of other SEZ villages over the total number of other villages in a county. $G_i$ is a binary variable equal to one if this ratio in county $c$ in which village $i$ is located is above the mean of the ratio among all counties.\footnote{Counties on average contain 148.3 villages. On average, there are 5.8 SEZ villages in a county.} In this way, exposure is characterized when there are sufficient number of SEZ villages in a county. It is in line with the initial goal of SEZ establishment, namely promoting agglomeration economies. Villages are considered as intensively exposed to neighbors' economic zones if the corresponding county contains a relatively high fraction of SEZ villages.

There are four baseline village characteristics including logs of a village’s distance from an airport and port, log of the capital-to-labor ratio, and log of the number of firms in the village in 2004. The distance from an airport and port can be considered as fixed attributes, while the capital-to-labor ratio and the number of firms can be treated as stochastic.\footnote{In the conditional inference framework adopted here, we can consider a subset of attributes fixed with the rest stochastic. The identification and estimation procedure stays the same. Only the inference will change in a way that the additional uncertainty resulting from a subset of stochastic attributes needs to be accounted for.} These baseline characteristics and their interactions with the direct treatment variable are included as regressors in moment conditions (\ref{eq:m3}) and (\ref{eq:m4}). The four baseline characteristics and their leave-one-out means at the county level are covariates in moment conditions (\ref{eq:m1}) and (\ref{eq:m2}) for the propensity scores. The spillover effects are estimated analogously using equations (\ref{eq:spillover1}) and (\ref{eq:spillover0}) in Appendix \ref{spilloversection}.

The first row of Table \ref{application} below reports DID estimates using the IPW approach in \cite{abadie2005semiparametric} with the four baseline village characteristics as covariates. Because of potential spillover effects, these canonical estimates are difficult to interpret causally. When interference has been considered, the direct eﬀects are mostly smaller than the canonical DID estimates, especially for SEZ villages with exposure level one. To summarize, SEZ establishment has positive and statistically significant direct effects at the 1\% level. This implies that the SEZ villages benefit from the program by gaining investment, employing more labor, and producing more output.

\begin{table}[htbp]
  \centering
  \caption{Direct and Spillover Effects of Special Economic Zones}
\begin{threeparttable}
    \begin{tabular}{lccc}
\toprule
          & log capital & log employment & log output  \\
\midrule
    Canonical DID & 0.689 & 0.395 & 0.629 \\
          & (0.049) & (0.038) & (0.051) \\
\midrule
 Direct effect (1) & 0.438 & 0.219 & 0.302 \\
         & (0.070) & (0.057) & (0.072) \\
   Direct effect (0) & 0.679 & 0.441 & 0.645 \\
          & (0.060) & (0.045) & (0.059) \\
    Spillover effect (treated) & -0.138 & -0.096 & -0.174 \\
          & (0.089) & (0.073) & (0.091) \\
    Spillover effect (untreated) & 0.103 & 0.129 & 0.169 \\
         & (0.033) & (0.030) & (0.042) \\
\bottomrule
    \end{tabular}%
 \begin{tablenotes}
   \begin{footnotesize}
    \item[1] The standard errors are clustered at the county level.
    \item[2] Canonical DID is estimated using inverse probability weighted DID in \cite{abadie2005semiparametric}.
   \item[3] Direct effect (1) and direct effect (0) are $\tau(1)$ and $\tau(0)$ respectively; spillover effect (treated) and spillover effect (untreated) are $\tau(1,1,0)$ and $\tau(0,1,0)$ as defined in Appendix \ref{spilloversection} respectively. 
\end{footnotesize}
\end{tablenotes}
\end{threeparttable}
  \label{application}%
\end{table}%

In terms of spillover effects, having relatively high ratio of neighboring SEZ villages does not significantly affect economic activities in SEZ villages. The only exception is for log output, where the spillover effect under more intensive exposure is quite negative and marginally statistically significant. By contrast, non-SEZ villages benefit from SEZ neighbors when there are sufficient number of SEZ villages in the same county. These patterns of direct and spillover effects are not found in \cite{lu2019place}.  

As suggested in Section \ref{sec:pre}, I also examined pre-trends with the exposure mapping as well as classical pre-trends for canonical DID estimation. Unfortunately, there is only one economic census period prior to treatment. As a result, the pre-trends are tested with China’s Annual Surveys of Industrial Firms (ASIF) used by \cite{lu2019place}, which contain more data in the pre-treatment period but only cover firms with relatively large sizes. I use ASIF data from 2004 and 2005. Given this is not the same dataset used for the main analysis, the results presented in Table \ref{table:pretrends} are only demonstrative. For canonical DID, only the differential pre-trend for log of employment is statistically significant at the 10\% level with small magnitude. None of the doubly robust estimates for placebo direct effects are significant.

\begin{table}[htbp]
  \centering
  \caption{Direct Effects of Special Economic Zones: Pre-trends}
\begin{threeparttable}
    \begin{tabular}{lccc}
\toprule
          & log capital & log employment & log output \\
\midrule
    Canonical DID & 0.000     & 0.051 & 0.039 \\
          & (0.039) & (0.029) & (0.045) \\
\midrule
    Direct effect (1) & -0.042 & 0.022 & 0.102 \\
          & (0.080) & (0.083) & (0.105) \\
    Direct effect (0) & -0.010 & 0.055 & -0.015 \\
          & (0.059) & (0.048) & (0.070) \\
\bottomrule
    \end{tabular}%
 \begin{tablenotes}
   \begin{footnotesize}
   \item[1] ASIF data from 2004 and 2005 are used in the pre-trends test. 
    \item[2] The standard errors are clustered at the county level.
    \item[3] Canonical DID is estimated using inverse probability weighted DID in \cite{abadie2005semiparametric}.
   \item[4] Direct effect (1) and direct effect (0) are $\tau(1)$ and $\tau(0)$ respectively. 
\end{footnotesize}
\end{tablenotes}
\end{threeparttable}
  \label{table:pretrends}%
\end{table}%

\section{Conclusion}
I propose doubly robust estimators for the direct treatment effect and spillover effect in a DID context. I later generalize to the case where exposure mapping could be misspecified and interference is not restricted within a fixed boundary of neighborhoods. Given the general spillover effect, one needs to account for spatial correlation when conducting inference. With the entire population observed, the usual spatial-correlation robust variance estimator could be conservative. 

I provide identification results of the direct and spillover effect for the IPW estimand, outcome regression estimand, and the doubly robust estimand. From here, researchers can approach these estimands using various parametric, semiparametric, or nonparametric estimation methods. In the current paper, I proved the asymptotic properties of GMM-type parametric estimators as an illustration of estimation. Given the inclusion of neighbors' treatments and attributes in the propensity score and the conditional mean functions, other nonparametric estimation methods are left as future work.

\bibliographystyle{agsm}

\bibliography{DID}

@article{forastiere2021identification,
  title={Identification and estimation of treatment and interference effects in observational studies on networks},
  author={Forastiere, Laura and Airoldi, Edoardo M and Mealli, Fabrizia},
  journal={Journal of the American Statistical Association},
  volume={116},
  number={534},
  pages={901--918},
  year={2021},
  publisher={Taylor \& Francis}
}

@techreport{jardim2022boundary,
  title={Boundary Discontinuity Methods and Policy Spillovers},
  author={Jardim, Ekaterina S and Long, Mark C and Plotnick, Robert and van Inwegen, Emma and Vigdor, Jacob L and Wething, Hilary},
  year={2022},
  institution={National Bureau of Economic Research}
}

@article{manski1993identification,
  title={Identification of endogenous social effects: The reflection problem},
  author={Manski, Charles F},
  journal={Review of Economic Studies},
  volume={60},
  number={3},
  pages={531--542},
  year={1993},
  publisher={Wiley-Blackwell}
}

@article{athey2022design,
  title={Design-based analysis in difference-in-differences settings with staggered adoption},
  author={Athey, Susan and Imbens, Guido W},
  journal={Journal of Econometrics},
  volume={226},
  number={1},
  pages={62--79},
  year={2022},
  publisher={Elsevier}
}

@techreport{rambachan2020design,
  title={Design-based uncertainty for quasi-experiments},
  author={Rambachan, Ashesh and Roth, Jonathan},
  Institution={arXiv preprint arXiv:2008.00602},
  year={2022}
}

@techreport{butts2021difference,
  title={Difference-in-differences estimation with spatial spillovers},
  author={Butts, Kyle},
  institution={arXiv preprint arXiv:2105.03737},
  year={2021}
}

@article{delgado2015difference,
  title={Difference-in-differences techniques for spatial data: Local autocorrelation and spatial interaction},
  author={Delgado, Michael S and Florax, Raymond JGM},
  journal={Economics Letters},
  volume={137},
  pages={123--126},
  year={2015},
  publisher={Elsevier}
}

@article{hudgens2008toward,
  title={Toward causal inference with interference},
  author={Hudgens, Michael G and Halloran, M Elizabeth},
  journal={Journal of the American Statistical Association},
  volume={103},
  number={482},
  pages={832--842},
  year={2008},
  publisher={Taylor \& Francis}
}

@article{aronow2017estimating,
  title={Estimating average causal effects under general interference, with application to a social network experiment},
  author={Aronow, Peter M and Samii, Cyrus},
  journal={Annals of Applied Statistics},
  volume={11},
  number={4},
  pages={1912--1947},
  year={2017},
  publisher={Institute of Mathematical Statistics}
}

@article{leung2022causal,
  title={Causal inference under approximate neighborhood interference},
  author={Leung, Michael P},
  journal={Econometrica},
  volume={90},
  number={1},
  pages={267--293},
  year={2022},
  publisher={Wiley Online Library}
}

@article{sant2020doubly,
  title={Doubly robust difference-in-differences estimators},
  author={Sant’Anna, Pedro HC and Zhao, Jun},
  journal={Journal of Econometrics},
  volume={219},
  number={1},
  pages={101--122},
  year={2020},
  publisher={Elsevier}
}

@article{savje2021average,
  title={Average treatment effects in the presence of unknown interference},
  author={S{\"a}vje, Fredrik and Aronow, Peter and Hudgens, Michael},
  journal={Annals of Statistics},
  volume={49},
  number={2},
  pages={673},
  year={2021},
  publisher={NIH Public Access}
}

@article{di2004police,
  title={Do police reduce crime? Estimates using the allocation of police forces after a terrorist attack},
  author={Di Tella, Rafael and Schargrodsky, Ernesto},
  journal={American Economic Review},
  volume={94},
  number={1},
  pages={115--133},
  year={2004},
  publisher={American Economic Association}
}

@article{manski2013identification,
  title={Identification of treatment response with social interactions},
  author={Manski, Charles F},
  journal={The Econometrics Journal},
  volume={16},
  number={1},
  pages={S1--S23},
  year={2013},
  publisher={Oxford University Press Oxford, UK}
}

@article{basse2018limitations,
  title={Limitations of design-based causal inference and A/B testing under arbitrary and network interference},
  author={Basse, Guillaume W and Airoldi, Edoardo M},
  journal={Sociological Methodology},
  volume={48},
  number={1},
  pages={136--151},
  year={2018},
  publisher={SAGE Publications Sage CA: Los Angeles, CA}
}

@article{kojevnikov2021limit,
  title={Limit theorems for network dependent random variables},
  author={Kojevnikov, Denis and Marmer, Vadim and Song, Kyungchul},
  journal={Journal of Econometrics},
  volume={222},
  number={2},
  pages={882--908},
  year={2021},
  publisher={Elsevier}
}

@article{jenish2009central,
  title={Central limit theorems and uniform laws of large numbers for arrays of random fields},
  author={Jenish, Nazgul and Prucha, Ingmar R},
  journal={Journal of Econometrics},
  volume={150},
  number={1},
  pages={86--98},
  year={2009},
  publisher={Elsevier}
}

@techreport{agarwal2022network,
  title={Network Synthetic Interventions: A Framework for Panel Data with Network Interference},
  author={Agarwal, Anish and Cen, Sarah and Shah, Devavrat and Yu, Christina Lee},
  institution={arXiv preprint arXiv:2210.11355},
  year={2022}
}

@techreport{emmenegger2022treatment,
  title={Treatment Effect Estimation from Observational Network Data using Augmented Inverse Probability Weighting and Machine Learning},
  author={Emmenegger, Corinne and Spohn, Meta-Lina and B{\"u}hlmann, Peter},
  institution={arXiv preprint arXiv:2206.14591},
  year={2022}
}

@techreport{qu2022efficient,
  title={Efficient treatment effect estimation in observational studies under heterogeneous partial interference},
  author={Qu, Zhaonan and Xiong, Ruoxuan and Liu, Jizhou and Imbens, Guido},
  institution={arXiv preprint arXiv:2107.12420},
  year={2022}
}

@techreport{auerbach2021local,
  title={The local approach to causal inference under network interference},
  author={Auerbach, Eric and Tabord-Meehan, Max},
  institution={arXiv preprint arXiv:2105.03810},
  year={2021}
}

@article{jin2023tailored,
  title={Tailored inference for finite populations: conditional validity and transfer across distributions},
  author={Jin, Ying and Rothenh{\"a}usler, Dominik},
  journal={Biometrika},
  volume={111},
  number={1},
  pages={215--233},
  year={2024},
  publisher={Oxford University Press}
}

@article{imai2014covariate,
  title={Covariate balancing propensity score},
  author={Imai, Kosuke and Ratkovic, Marc},
  journal={Journal of the Royal Statistical Society: Series B: Statistical Methodology},
  pages={243--263},
  year={2014},
  publisher={JSTOR}
}

@article{jenish2012spatial,
  title={On spatial processes and asymptotic inference under near-epoch dependence},
  author={Jenish, Nazgul and Prucha, Ingmar R},
  journal={Journal of Econometrics},
  volume={170},
  number={1},
  pages={178--190},
  year={2012},
  publisher={Elsevier}
}

@book{gallant1988unified,
  title={A unified theory of estimation and inference for nonlinear dynamic models},
  author={Gallant, A Ronald and White, Halbert},
  year={1988},
  publisher={Blackwell}
}

@article{newey1994large,
	Author = {Newey, Whitney K and McFadden, Daniel},
	Journal = {Handbook of Econometrics},
	Pages = {2111--2245},
	Publisher = {Elsevier},
	Title = {Large sample estimation and hypothesis testing},
	Volume = {4},
	Year = {1994}}

@techreport{xu2022design,
  title={A Design-Based Approach to Spatial Correlation},
  author={Xu, Ruonan and Wooldridge, Jeffrey M},
  institution={arXiv preprint arXiv:2211.14354},
  year={2022}
}

@article{newey1991uniform,
  title={Uniform convergence in probability and stochastic equicontinuity},
  author={Newey, Whitney K},
  journal={Econometrica},
  pages={1161--1167},
  year={1991},
  volume={59},
  publisher={JSTOR}
}

@article{abadie2014inference,
  title={Inference for misspecified models with fixed regressors},
  author={Abadie, Alberto and Imbens, Guido W and Zheng, Fanyin},
  journal={Journal of the American Statistical Association},
  volume={109},
  number={508},
  pages={1601--1614},
  year={2014},
  publisher={Taylor \& Francis}
}

@article{abadie2005semiparametric,
  title={Semiparametric difference-in-differences estimators},
  author={Abadie, Alberto},
  journal={Review of Economic Studies},
  volume={72},
  number={1},
  pages={1--19},
  year={2005},
  publisher={Wiley-Blackwell}
}

@techreport{man2023doubly,
  title={Doubly Robust Estimators with Weak Overlap},
  author={Man, Yukun and Sant'Anna, Pedro HC and Sasaki, Yuya and Ura, Takuya},
  institution={arXiv preprint arXiv:2304.08974},
  year={2023}
}

@article{roth2023parallel,
  title={When is parallel trends sensitive to functional form?},
  author={Roth, Jonathan and Sant'Anna, Pedro HC},
  journal={Econometrica},
  volume={91},
  number={2},
  pages={737--747},
  year={2023},
  publisher={Wiley Online Library}
}

@techreport{ghanem2022selection,
  title={Selection and parallel trends},
  author={Ghanem, Dalia and Sant'Anna, Pedro HC and W{\"u}thrich, Kaspar},
  institution={arXiv preprint arXiv:2203.09001},
  year={2022}
}

@article{rambachan2023more,
  title={A more credible approach to parallel trends},
  author={Rambachan, Ashesh and Roth, Jonathan},
  journal={Review of Economic Studies},
  volume={90},
  number={5},
  pages={2555--2591},
  year={2023},
  publisher={Oxford University Press US}
}

@article{roth2022pretest,
  title={Pretest with caution: Event-study estimates after testing for parallel trends},
  author={Roth, Jonathan},
  journal={American Economic Review: Insights},
  volume={4},
  number={3},
  pages={305--22},
  year={2022}
}

@article{savje2023causal,
	title={Causal inference with misspecified exposure mappings: separating definitions and assumptions},
	author={S{\"a}vje, Fredrik},
	journal={Biometrika},
	volume={111},
	number={1},
	pages={1--15},
	year={2024},
	publisher={Oxford University Press}
}

@article{ma2020robust,
  title={Robust inference using inverse probability weighting},
  author={Ma, Xinwei and Wang, Jingshen},
  journal={Journal of the American Statistical Association},
  volume={115},
  number={532},
  pages={1851--1860},
  year={2020},
  publisher={Taylor \& Francis}
}

@techreport{abadie2002simple,
  title={Simple and bias-corrected matching estimators for average treatment effects},
  author={Abadie, Alberto and Imbens, Guido},
  year={2002},
  institution={National Bureau of Economic Research}
}

@techreport{balzer2015targeted,
  title={Targeted estimation and inference for the sample average treatment effect},
  author={Balzer, Laura B and Petersen, Maya L and van der Laan, Mark J},
  year={2015},
  institution={U.C. Berkeley Division of Biostatistics Working Paper Series}
}

@article{imbens2004nonparametric,
  title={Nonparametric estimation of average treatment effects under exogeneity: A review},
  author={Imbens, Guido W},
  journal={Review of Economics and statistics},
  volume={86},
  number={1},
  pages={4--29},
  year={2004},
  publisher={MIT Press 238 Main St., Suite 500, Cambridge, MA 02142-1046, USA journals~…}
}

@article{huber2021framework,
  title={A framework for separating individual-level treatment effects from spillover effects},
  author={Huber, Martin and Steinmayr, Andreas},
  journal={Journal of Business \& Economic Statistics},
  volume={39},
  number={2},
  pages={422--436},
  year={2021},
  publisher={Taylor \& Francis}
}

@article{lu2019place,
  title={Place-based policies, creation, and agglomeration economies: Evidence from China’s economic zone program},
  author={Lu, Yi and Wang, Jin and Zhu, Lianming},
  journal={American Economic Journal: Economic Policy},
  volume={11},
  number={3},
  pages={325--360},
  year={2019},
  publisher={American Economic Association}
}

@article{viviano2024policy,
  title={Policy targeting under network interference},
  author={Viviano, Davide},
  journal={Review of Economic Studies},
  year={2024}
}

@article{graham2012inverse,
  title={Inverse probability tilting for moment condition models with missing data},
  author={Graham, Bryan S and de Xavier Pinto, Cristine Campos and Egel, Daniel},
  journal={Review of Economic Studies},
  volume={79},
  number={3},
  pages={1053--1079},
  year={2012},
  publisher={Oxford University Press}
}

\appendix
\numberwithin{equation}{section}
\numberwithin{assumption}{section}

\section{Proofs}\label{sec:proof}

\begin{definition}
The random function $g(X_i,\theta)$ is said to be Lipschitz in parameter $\theta$ on $\Theta$ if there is $h(u)\downarrow 0$ as $u\downarrow 0$ and $b(\cdot): \mathcal{W}\to R$ such that $\sup _{M, i\in D_M}\mathbb{E}\big[|b(X_i)|\big]<\infty$, and for all $\tilde{\theta},\theta\in\Theta$, $\big|g(X_i,\tilde{\theta})-g(X_i,\theta)\big|\leq b(X_i) h(\|\tilde{\theta}-\theta\|)$, $i\in D_M, M\geq 1$.
\end{definition}

\begin{assumption}\label{ap:gmm}
(\romannumeral 1) $\widehat{\Psi}-\Psi_M\overset{p}\to \textbf{0}$, where $\Psi_M$ is positive semidefinite; \\
(\romannumeral 2) $\Theta$ is compact; \\
(\romannumeral 3) let $Q_M(\theta) =\mathbbm{E}_D\big[q(X_i,\theta)\big]'\Psi\mathbbm{E}_D\big[q(X_i,\theta)\big]$. $\{Q_M(\theta)\}$ has identifiably unique minimizers $\{\theta_M^*\}$ on $\Theta$ as in Definition 3.2 in \cite{gallant1988unified}; \\
(\romannumeral 4) $q(X_i,\theta)$ is continuously differentiable on $int(\Theta)$, $\forall\ i,M$;  \\
(\romannumeral 5) $q(X_i,\theta)$ is Lipschitz in $\theta$ on $\Theta$;\\
(\romannumeral 6) $\sup _{M, i\in D_M}\mathbb{E}\Big[\sup _{\theta\in\Theta}\norm{q(X_i,\theta)}^p\big| \bm{z}\Big]<\infty$ for some $p>4$;\\
(\romannumeral 7) $\theta^*_M\in int(\Theta)$ uniformly in $M$, and $\mathbbm{E}_D\big[q(X_i,\theta^*_M)\big]=\bf{0}$;\\
(\romannumeral 8) $\inf_M\lambda_{min}(\Omega_M)>0$, where $\lambda_{min}(\cdot)$ is the smallest eigenvalue;\\
(\romannumeral 9) $\nabla_\theta q(X_i,\theta)$ is Lipschitz in $\theta$ on $\Theta$;\\
(\romannumeral 10) $\sup _{M, i\in D_M}\mathbb{E}\Big[\sup _{\theta\in\Theta}\norm{\nabla_\theta q(X_i,\theta)}^2\big|\bm{z}\Big]<\infty$;\\
(\romannumeral 11) ${R_M^*}'\Psi_M R_M^*$ is nonsingular;\\
(\romannumeral 12) let $l_i=l(X_i,\theta)$ be a generic function standing for each element of either $q(X_i,\theta)$ or $\nabla_\theta q(X_i,\theta)$. $\forall\ \theta\in\Theta$, $l(X_i,\theta)$ is Lipschitz in $X_i$ on the domain of $X_i$ such that $\sup_{M,i\in D_M}\text{Lip}(l_i)<\infty$.
\end{assumption}

Notice that a necessary condition for Assumption \ref{ap:gmm}(\romannumeral 12) is
$\sup_{M,i\in D_M}|Y_{it}|\leq C<\infty$ and $\sup_{M,i\in D_M}\norm{z_i}\leq C<\infty$, which can often imply Assumption \ref{ap:gmm}(\romannumeral 6) and (\romannumeral 10).

\bigskip
\noindent
\textbf{Proof of Proposition \ref{pro1}}

Compare the canonical DID estimand with the overall direct effect: 
\begin{align*}
	\tau=&\frac{1}{|D_M|}\sum_{i\in D_M}\mathbbm{E}\big[y_{i2}(1,\bm{W}_{-i})-y_{i2}(0,\bm{W}_{-i})|W_i=1, z_i\big]\\
	=&\frac{1}{|D_M|}\sum_{i\in D_M}\sum_{g\in \mathcal{G}}\mathbbm{E}\big[y_{i2}(1,\bm{W}_{-i})-y_{i2}(0,\bm{W}_{-i})|W_i=1, G_i=g, z_i\big] P(G_i=g|W_i=1,z_i)\\
	=&\frac{1}{|D_M|}\sum_{i\in D_M}\sum_{g\in \mathcal{G}}\mathbbm{E}\big[\tilde{y}_{i2}(1,g)-\tilde{y}_{i2}(0,g)|W_i=1, G_i=g, z_i\big] P(G_i=g|W_i=1,z_i)\\
	=&\frac{1}{|D_M|}\sum_{i\in D_M}\sum_{g\in \mathcal{G}}\Big\{\mathbbm{E}\big(\tilde{y}_{i2}(1,g)|W_i=1, G_i=g, z_i\big)-\mathbbm{E}\big(y_{i1}(0,\underline{0})|W_i=1,G_i=g,z_i\big)\\
	&-\Big[\mathbbm{E}\big(\tilde{y}_{i2}(0, g)|W_i=0,G_i=g, z_i\big)-\mathbbm{E}\big(y_{i1}(0,\underline{0})|W_i=0,G_i=g,z_i\big)\Big]\Big\} P(G_i=g|W_i=1,z_i)\\
	=&\frac{1}{|D_M|}\sum_{i\in D_M}\bigg\{\sum_{g\in \mathcal{G}}\Big[\mathbbm{E}(Y_{i2}|W_i=1,G_i=g,z_i)-\mathbbm{E}(Y_{i1}|W_i=1,G_i=g,z_i)\Big]P(G_i=g|W_i=1,z_i)\\
	&-\sum_{g\in \mathcal{G}}\Big[\mathbbm{E}(Y_{i2}|W_i=0,G_i=g,z_i)-\mathbbm{E}(Y_{i1}|W_i=0,G_i=g,z_i)\Big]P(G_i=g|W_i=1,z_i)\bigg\}
\numberthis \label{eq:p1}
\end{align*}

\begin{align*}
	\tau_{canonic}=&\frac{1}{|D_M|}\sum_{i\in D_M}\Big[\mathbbm{E}(Y_{i2}-Y_{i1}|W_i=1,z_i)-\mathbbm{E}(Y_{i2}-Y_{i1}|W_i=0,z_i)\Big]\\
	=&\frac{1}{|D_M|}\sum_{i\in D_M}\bigg\{\sum_{g\in \mathcal{G}}\Big[\mathbbm{E}(Y_{i2}|W_i=1,G_i=g,z_i)-\mathbbm{E}(Y_{i1}|W_i=1,G_i=g,z_i)\Big]P(G_i=g|W_i=1,z_i)\\
	&-\sum_{g\in \mathcal{G}}\Big[\mathbbm{E}(Y_{i2}|W_i=0,G_i=g,z_i)-\mathbbm{E}(Y_{i1}|W_i=0,G_i=g,z_i)\Big]P(G_i=g|W_i=0,z_i)\bigg\}
	\tag{\stepcounter{equation}\theequation}
\end{align*}

\noindent\textbf{Proof of Proposition \ref{pro3}}

Identification of the doubly robust estimand:

When the propensity scores are correctly specified, $\eta(z)=p(z)$, $\eta_{1g}(z)=\pi_{1g}(z)$, and $\eta_{0g}(z)=\pi_{0g}(z)$.
\begin{align*}
	&\mathbbm{E}\Bigg[\frac{W_i}{p(z_i)}\frac{\mathbbm{1}\{G_i=g\}}{\pi_{1g}(z_i)} \Big(\big(Y_{i2}-m_{2,1g}(z_i)\big)-\big(Y_{i1}-m_{1,1g}(z_i)\big)\Big)\bigg|z_i\Bigg]\\
	=&\mathbbm{E}\Bigg[\frac{W_i}{p(z_i)} \frac{\mathbbm{1}\{G_i=g\}}{\pi_{1g}(z_i)} \Big(\big(Y_{i2}-m_{2,1g}(z_i)\big)-\big(Y_{i1}-m_{1,1g}(z_i)\big)\Big)\bigg|z_i,W_i=1\Bigg]P(W_i=1|z_i)\\
	=&\mathbbm{E}\bigg[\frac{\mathbbm{1}\{G_i=g\}}{\pi_{1g}(z_i)} \Big(\big(Y_{i2}-m_{2,1g}(z_i)\big)-\big(Y_{i1}-m_{1,1g}(z_i)\big)\Big)\bigg|z_i,W_i=1,G_i=g\bigg]P(G_i=g|W_i=1,z_i)\\
	=&\mathbbm{E}(Y_{i2}|z_i,W_i=1,G_i=g)-\mathbbm{E}(Y_{i1}|z_i,W_i=1,G_i=g)-\big[m_{2,1g}(z_i)-m_{1,1g}(z_i)\big]
	\tag{\stepcounter{equation}\theequation}
\end{align*}
Similarly,
\begin{equation}
	\begin{aligned}
		&\mathbbm{E}\Bigg[\frac{1-W_i}{1-p(z_i)}\frac{\mathbbm{1}\{G_i=g\}}{\pi_{0g}(z_i)} \Big(\big(Y_{i2}-m_{2,0g}(z_i)\big)-\big(Y_{i1}-m_{1,0g}(z_i)\big)\Big)\bigg|z_i\Bigg]\\
		=&\mathbbm{E}(Y_{i2}|z_i,W_i=0,G_i=g)-\mathbbm{E}(Y_{i1}|z_i,W_i=0,G_i=g)-\big[m_{2,0g}(z_i)-m_{1,0g}(z_i)\big]
	\end{aligned}
\end{equation}
Hence, 
\begin{align*}
	&\mathbbm{E}_D\Bigg[\frac{W_i}{p(z_i)}\frac{\mathbbm{1}\{G_i=g\}}{\pi_{1g}(z_i)} \Big(\big(Y_{i2}-m_{2,1g}(z_i)\big)-\big(Y_{i1}-m_{1,1g}(z_i)\big)\Big)\\
	&-\frac{1-W_i}{1-p(z_i)} \frac{\mathbbm{1}\{G_i=g\}}{\pi_{0g}(z_i)}\Big(\big(Y_{i2}-m_{2,0g}(z_i)\big)-\big(Y_{i1}-m_{1,0g}(z_i)\big)\Big)+\Delta m_{1g}(z_i)-\Delta m_{0g}(z_i)\Bigg]\\
	=&\frac{1}{|D_M|}\sum_{i\in D_M}\bigg\{\mathbbm{E}(Y_{i2}|z_i,W_i=1,G_i=g)-\mathbbm{E}(Y_{i1}|z_i,W_i=1,G_i=g)\\
	&-\big[\mathbbm{E}(Y_{i2}|z_i,W_i=0,G_i=g)-\mathbbm{E}(Y_{i1}|z_i,W_i=0,G_i=g)\big]\\
	&-\Big(\big[m_{2,1g}(z_i)-m_{1,1g}(z_i)\big]-\big[m_{2,0g}(z_i)-m_{1,0g}(z_i)\big]\Big)+\Delta m_{1g}(z_i)-\Delta m_{0g}(z_i)\bigg\}\\
	=&\frac{1}{|D_M|}\sum_{i\in D_M}\mathbbm{E}\big[\tilde{y}_{i2}(1,g)-\tilde{y}_{i2}(0,g)|W_i=1, G_i=g, z_i\big]
	\tag{\stepcounter{equation}\theequation}
\end{align*}

When the conditional means are correctly specified, $\Delta m_{wg}(z)=\mu_{2,wg}(z)-\mu_{1,wg}(z)\equiv \Delta \mu_{wg}(z)$ for $w\in\{0,1\}$.
\begin{align*}
	&\mathbbm{E}\Bigg[\frac{W_i}{\eta(z_i)} \frac{\mathbbm{1}\{G_i=g\}}{\eta_{1g}(z_i)}\Big(\big(Y_{i2}-\mu_{2,1g}(z_i)\big)-\big(Y_{i1}-\mu_{1,1g}(z_i)\big)\Big)\bigg|z_i\Bigg]\\
	=&\mathbbm{E}\Bigg[\frac{W_i}{\eta(z_i)}\frac{\mathbbm{1}\{G_i=g\}}{\eta_{1g}(z_i)} \Big(\big(Y_{i2}-\mu_{2,1g}(z_i)\big)-\big(Y_{i1}-\mu_{1,1g}(z_i)\big)\Big)\bigg|z_i,W_i=1\Bigg]P(W_i=1|z_i)\\
	=&\frac{p(z_i)}{\eta(z_i)}\mathbbm{E}\bigg[ \frac{\mathbbm{1}\{G_i=g\}}{\eta_{1g}(z_i)}\Big(\big(Y_{i2}-\mu_{2,1g}(z_i)\big)-\big(Y_{i1}-\mu_{1,1g}(z_i)\big)\Big)\bigg|z_i,W_i=1,G_i=g\bigg]P(G_i=g|z_i,W_i=1)\\
	=&\frac{p(z_i)}{\eta(z_i)}\frac{\pi_{1g}(z_i)}{\eta_{1g}(z_i)}\Big[\mathbbm{E}(Y_{i2}|z_i,W_i=1,G_i=g)-\mu_{2,1g}(z_i)-\big(\mathbbm{E}(Y_{i1}|z_i,W_i=1,G_i=g)-\mu_{1,1g}(z_i)\big)\Big]=0
	\tag{\stepcounter{equation}\theequation}
\end{align*}
Analogously, 
\begin{equation}
	\begin{aligned}
		\mathbbm{E}\Bigg[\frac{1-W_i}{1-\eta(z_i)}\frac{\mathbbm{1}\{G_i=g\}}{\eta_{0g}(z_i)} \Big(\big(Y_{i2}-\mu_{2,0g}(z_i)\big)-\big(Y_{i1}-\mu_{1,0g}(z_i)\big)\Big)\bigg|z_i\Bigg]=0
	\end{aligned}
\end{equation}
As a result, 
\begin{align*}
	&\mathbbm{E}_D\Bigg[\frac{W_i}{\eta(z_i)}\frac{\mathbbm{1}\{G_i=g\}}{\eta_{1g}(z_i)} \Big(\big(Y_{i2}-\mu_{2,1g}(z_i)\big)-\big(Y_{i1}-\mu_{1,1g}(z_i)\big)\Big)\\
	&-\frac{1-W_i}{1-\eta(z_i)}\frac{\mathbbm{1}\{G_i=g\}}{\eta_{0g}(z_i)} \Big(\big(Y_{i2}-\mu_{2,0g}(z_i)\big)-\big(Y_{i1}-\mu_{1,0g}(z_i)\big)\Big)+\Delta \mu_{1g}(z_i)-\Delta \mu_{0g}(z_i)\Bigg]\\
	=&\frac{1}{|D_M|}\sum_{i\in D_M}\big[\Delta \mu_{1g}(z_i)-\Delta \mu_{0g}(z_i)\big]\\
	=&\frac{1}{|D_M|}\sum_{i\in D_M}\mathbbm{E}\big[\tilde{y}_{i2}(1,g)-\tilde{y}_{i2}(0,g)|W_i=1, G_i=g, z_i\big]
	\tag{\stepcounter{equation}\theequation}
\end{align*}

Proof for Corollary \ref{coro1} is exactly the same and hence is omitted here.

\bigskip
\noindent
\textbf{Proof of Lemma \ref{weakdependence}}: 

Denote $l_i^{(r)}=l\big(X_i^{(r)},\theta\big)=l\big(y_{it}\big(\bm{W}^{(i,r,0)}\big),G\big(i,\bm{W}^{(i,r,0)}_{-i}\big),W_i, z_i,\theta\big)$.
Let $f\in \mathcal{L}_{\nu,h}$ and $f'\in \mathcal{L}_{\nu,h'}$. Let $s>0$ and $(H,H')\in \mathcal{P}_M(h,h';s)$. Define $\xi=f(l_H)$, $\zeta=f'(l_{H'})$, $\xi^{(s)}=f(l_i^{(s)}: i\in H)$, and $\zeta^{(s)}=f'(l_i^{(s)}: i\in H')$.

First, for $s\leq 3\max\{K,\rho_0\}$, we have 
\begin{align*}
	\big|Cov(\xi,\zeta|\bm{z})\big|\leq 2 \norm{f}_\infty\norm{f'}_\infty\leq C_1<\infty
	\tag{\stepcounter{equation}\theequation}
\end{align*}
Next, consider $s>3\max\{K,\rho_0\}$. 
\begin{equation}\label{eq:weak}
	\begin{aligned}
		&\big|Cov(\xi,\zeta|\bm{z})\big|=\big|Cov(\xi-\xi^{(s/3)}+\xi^{(s/3)},\zeta|\bm{z})\big|\\
		\leq&\big|Cov(\xi-\xi^{(s/3)},\zeta|\bm{z})\big|+\big|Cov(\xi^{(s/3)},\zeta-\zeta^{(s/3)}|\bm{z})\big|+\big|Cov(\xi^{(s/3)},\zeta^{(s/3)}|\bm{z})\big|\\
		\leq& 2\norm{f'}_\infty\mathbbm{E}\Big[\big|\xi-\xi^{(s/3)}\big|\Big|\bm{z}\Big]+2\norm{f}_\infty\mathbbm{E}\Big[\big|\zeta-\zeta^{(s/3)}\big|\Big|\bm{z}\Big]+\Big|Cov\big(\xi^{(s/3)},\zeta^{(s/3)}|\bm{z}\big)\Big|
	\end{aligned}
\end{equation}
For the first two terms in equation (\ref{eq:weak}), 
\begin{align*}
	&\norm{f'}_\infty\mathbbm{E}\Big[\big|\xi-\xi^{(s/3)}\big|\Big|\bm{z}\Big]+\norm{f}_\infty\mathbbm{E}\Big[\big|\zeta-\zeta^{(s/3)}\big|\Big|\bm{z}\Big]\\
	\leq & h\norm{f'}_\infty \text{Lip}(f)\sup_{M,i\in D_M}\mathbbm{E}\Big[\big|l_i-l_i^{(s/3)}\big|\Big|\bm{z}\Big]+h'\norm{f}_\infty \text{Lip}(f')\sup_{M,i\in D_M}\mathbbm{E}\Big[\big|l_i-l_i^{(s/3)}\big|\Big|\bm{z}\Big]\\
	\leq &\big[h\norm{f'}_\infty \text{Lip}(f)+h'\norm{f}_\infty \text{Lip}(f')\big]\sup_{M,i\in D_M}\text{Lip}(l_i)\sup_{M,i\in D_M}\mathbbm{E}\Big[\big\|X_i-X_i^{(s/3)}\big\|\Big|\bm{z}\Big]
	\tag{\stepcounter{equation}\theequation}
\end{align*}
Since $s/3\geq K$,
\[
\big(Y_{i1}, y_{i2}\big(\bm{W}^{(i,s/3,0)}\big),G\big(i,\bm{W}^{(i,s/3,0)}_{-i}\big),W_i, z_i\big)=\big(Y_{i1}, y_{i2}\big(\bm{W}^{(i,s/3,0)}\big),G\big(i,\bm{W}_{-i}\big),W_i, z_i\big).
\]
As a result, 
\begin{equation}
	\mathbbm{E}\Big[\big\|X_i-X_i^{(s/3)}\big\|\Big|\bm{z}\Big]=\mathbbm{E}\Big[\big|y_{i2}(\bm{W})-y_{i2}\big(\bm{W}^{(i,s/3,0)}\big)\big|\Big|\bm{z}\Big]\leq \kappa_M(s/3).
\end{equation}

For any fixed $s$, $l_i^{(s/3)}$ is $\alpha$-mixing under Assumption \ref{ap:mixing}. By Proposition 2.2 in \cite{kojevnikov2021limit}, the last term in equation (\ref{eq:weak}) is bounded by 
\begin{equation}
	C_2\alpha^{l^{(s/3)}}(h,h',s)\leq C_2\alpha^\epsilon\Big(h C_3 \big(\frac{s}{3}\big)^d, h' C_3 \big(\frac{s}{3}\big)^d, \frac{s}{3}\Big).
\end{equation}
Putting these together, equation (\ref{eq:weak}) is bounded by 
\begin{equation}
	C\big(\kappa_M(s/3)+s^d\widehat{\alpha}^{\epsilon}(s/3)\big).
\end{equation} 

When the exposure mapping is correctly specified, Assumption 5 holds by definition for any $s\geq K$. Hence, Lemma 4.1 holds with $\tilde{\kappa}^*_{M,s}=s^d\widehat{\alpha}^{\epsilon}(s/3)\mathbbm{1}(s> 3\max\{K,\rho_0\})+\mathbbm{1}(s\leq 3\max\{K,\rho_0\})\leq \tilde{\kappa}_{M,s}$. Assumptions 7-9 are satisfied with $\tilde{\kappa}^*_{M,s}$.

\bigskip

\noindent\textbf{Proof of Theorem \ref{thm:clt}}: 

I prove the theorem by verifying Theorem 2.1 and Theorem 3.2 in \cite{newey1994large}. I first show $\hat{\theta}-\theta^*_M\overset{p}\to \bm{0}$. 

Under Assumption \ref{ap:gmm}(\romannumeral 6) and Assumption \ref{ap:wlln}
\begin{equation}\label{eq:qlln}
	\frac{1}{|D_M|}\sum_{i\in D_M} q(X_i,\theta)-\mathbbm{E}_D\big[q(X_i,\theta)\big]\overset{p}\to \bm{0}
\end{equation}
follows from Lemma \ref{weakdependence} and Theorem 3.1 in \cite{kojevnikov2021limit}.  
Next, 
\begin{equation}
	\sup_{\theta\in \Theta}\norm{\frac{1}{|D_M|}\sum_{i\in D_M} q(X_i,\theta)-\mathbbm{E}_D\big[q(X_i,\theta)\big]}\overset{p}\to \bm{0}
\end{equation} 
follows from Corollary 3.1 in \cite{newey1991uniform} and equation (\ref{eq:qlln}) under condition (\romannumeral 5). Also, $\mathbbm{E}_D\big[q(X_i,\theta)\big]$ is uniformly equicontinuous. Let 
\[
\widehat{Q}(\theta)=\frac{1}{|D_M|}\sum_{i\in D_M}q(X_i,\theta)'\widehat{\Psi}\frac{1}{|D_M|}\sum_{i\in D_M}q(X_i,\theta).
\]
Finally, I need to show
\begin{equation}\label{eq:qulln}
	\sup_{\theta\in\Theta}|\widehat{Q}(\theta)-Q_M(\theta)|\overset{p}\to 0
\end{equation} 
and $Q_M(\theta)$ is uniformly equicontinuous. The proof of equation (\ref{eq:qulln}) and the equicontinuity is standard. One can follow, for instance, the proof of Theorem 3 in \cite{jenish2012spatial}. 

Next, I prove the asymptotic normality. The key steps are to prove 
\begin{equation}\label{eq:clt}
	\Omega_M^{-1/2}\frac{1}{\sqrt{|D_M|}}\sum_{i\in D_M}q(X_i,\theta_M^*)\overset{d}\to \mathcal{N}(\textbf{0},I_k)
\end{equation} 
and 
\begin{equation}\label{eq:hulln}
	\sup_{\theta\in\Theta}\norm{\frac{1}{|D_M|}\sum_{i\in D_M}\nabla_\theta q(X_i,\theta)-\mathbbm{E}_D\big[\nabla_\theta q(X_i,\theta)\big]}\overset{p}\to \textbf{0}.
\end{equation}

Equation (\ref{eq:clt}) is implied by Theorem 3.2 in \cite{kojevnikov2021limit}, Lemma \ref{weakdependence}, and the Cramer-Wold device under Assumption \ref{ap:gmm}(\romannumeral 6) and (\romannumeral 8) and Assumption \ref{ap:clt}. By analogous argumentation for the proof of consistency, equation (\ref{eq:hulln}) holds under Assumption \ref{ap:gmm}(\romannumeral 9) and (\romannumeral 10).

\bigskip

\noindent\textbf{Proof of Theorem 4.3}: 

Using analogous arguments in the proof of Theorem 4.2, $\hat{R}-R_M^*\overset{p}\to \bm{0}$.  
The key step is to show that $\tilde{\Omega}(\hat{\theta})-\Omega_M-\Omega_E\overset{p}\to \bm{0}$.

Notice that 
\begin{align*}
	\Omega_M=&\frac{1}{|D_M|}\sum_{i\in D_M}\sum_{j\in D_M}\mathbb{E}\bigg\{\Big(q(X_i,\theta^*_M)-\mathbb{E}\big[q(X_i,\theta^*_M)|\bm{z}\big]\Big)\cdot\Big(q(X_j,\theta^*_M)-\mathbb{E}\big[q(X_j,\theta^*_M)|\bm{z}\big]\Big)'\Big|\bm{z}\bigg\}\\
	=&\frac{1}{|D_M|}\sum_{i\in D_M}\sum_{j\in D_M}\mathbb{E}\big(\tilde{q}(X_i,\theta^*_M)\tilde{q}(X_j,\theta^*_M)'\big),
	\tag{\stepcounter{equation}\theequation}
\end{align*}
where 
\begin{equation}
	\tilde{q}(X_i,\theta^*_M)=q(X_i,\theta^*_M)-\mathbb{E}\big[q(X_i,\theta^*_M)|\bm{z}\big]
\end{equation}
with $\mathbbm{E}\big[\tilde{q}(X_i,\theta^*_M)|\bm{z}\big]=\bm{0}$.

Since any sequence of symmetric matrices $\{A_N\}$ converges to a symmetric matrix $\{A_0\}$ if and only if $c'A_Nc\to c'A_0c$ for any vectors c, we can reach our conclusion by taking an arbitrary linear combination of $q(X_i,\theta)$. From now on, I focus on the case of a scalar $q(X_i,\theta)$.

\begin{equation}\label{eq14}
	\begin{aligned}
		\norm{\tilde{\Omega}(\hat{\theta})-\Omega_M-\Omega_E}\leq \norm{\tilde{\Omega}(\hat{\theta})-\tilde{\Omega}(\theta^*_M)}+\norm{\tilde{\Omega}(\theta^*_M)-\Omega_M-\Omega_E}.
	\end{aligned}
\end{equation}

For the first term on the right hand side of (\ref{eq14}), take a mean value expansion of $\tilde{\Omega}(\hat{\theta})$ around $\theta^*_M$. Let $\check{\theta}$ denote the mean value from this expansion. 
\begin{align*}
	&|\tilde{\Omega}(\hat{\theta})-\tilde{\Omega}(\theta^*_M)|\\
	=&\Bigg|(\hat{\theta}-\theta^*_M)\frac{1}{|D_M|}\sum^{\infty}_{s=0}\omega\Big(\frac{s}{b_M}\Big)\sum_{i\in D_M}\sum_{j\in D_M, s\leq \rho(i,j)<s+1}\big[\nabla_\theta q(X_i,\check{\theta})q(X_j,\check{\theta})+ q(X_j,\check{\theta})\nabla_\theta q(X_j,\check{\theta})\big]\Bigg|\\
	\leq & C_1 \big|\sqrt{|D_M|}(\hat{\theta}-\theta^*_M)\big|\frac{1}{|D_M|^{3/2}}\sum^{b_M}_{s=0}\sum_{i\in D_M}\sum_{j\in D_M, s\leq \rho(i,j)<s+1}\sup_{\theta\in\Theta}\big|\nabla_\theta q(X_i,\theta) q(X_j,\theta)\big|\\
	\leq& C \big|\sqrt{|D_M|}(\hat{\theta}-\theta^*_M)\big|\frac{1}{\sqrt{|D_M|}}\bigg(\sum^{b_M}_{s=1}s^{d-1}+1\bigg)\frac{1}{|D_M|}\sum_{i\in D_M}\sup_{\theta\in\Theta}\big|\nabla_\theta q(X_i,\theta) q(X_j,\theta)\big|
	\tag{\stepcounter{equation}\theequation}
\end{align*}
Since 
\begin{align*}
	&\mathbb{E}\Big[\frac{1}{|D_M|}\sum_{i\in D_M}\sup_{\theta\in\Theta}\big|\nabla_\theta q(X_i,\theta) q(X_j,\theta)\big|\Big|\bm{z}\Big]\leq \sup_{M,i\in D_M}\mathbb{E}\Big[\sup_{\theta\in\Theta}\big|\nabla_\theta q(X_i,\theta) q(X_j,\theta)\big|\Big|\bm{z}\Big]\\
	\leq & \sup_{M,i\in D_M}\mathbb{E}\Big[\sup_{\theta\in\Theta}\big|\nabla_\theta q(X_i,\theta)\big|^2\Big|\bm{z}\Big]^{1/2}\cdot \sup_{M,i\in D_M}\mathbb{E}\Big[\sup_{\theta\in\Theta}\big| q(X_i,\theta)\big|^2\Big|\bm{z}\Big]^{1/2}<\infty,
	\tag{\stepcounter{equation}\theequation}
\end{align*}
\begin{equation}
	\frac{1}{|D_M|}\sum_{i\in D_M}\sup_{\theta\in\Theta}\big|\nabla_\theta q(X_i,\theta) q(X_j,\theta)\big|=O_p(1)
\end{equation}
by Markov's inequality. 
Given $b_M=o\big(|D_M|^{1/2d}\big)$, 
$\frac{1}{\sqrt{|D_M|}}\sum_{s=1}^{b_M} s^{d-1}=o(1)$.
Also, $\sqrt{|D_M|}(\hat{\theta}-\theta^*_M)=O_p(1)$ by Theorem 4.2. Hence, $|\tilde{\Omega}(\hat{\theta})-\tilde{\Omega}(\theta^*_M)|=o_p(1)$.

Let
\begin{equation}
	\check{\Omega}_M=\frac{1}{|D_M|}\sum_{s=0}^{\infty}\omega\Big(\frac{s}{b_M}\Big)\sum_{i\in D_M}\sum_{j\in D_M, s\leq \rho(i,j)< s+1}\tilde{q}(X_i,\theta^*_M)\tilde{q}(X_j,\theta^*_M).
\end{equation}
Applying Proposition 4.1 in \cite{kojevnikov2021limit}, we have 
\begin{equation}
	\norm{\check{\Omega}_M-\Omega_M}=o_p(1).
\end{equation}
What is left is to show 
\begin{equation}\label{eq:last}
	\begin{aligned}
		&\norm{\tilde{\Omega}(\theta^*_M)-\Omega_E-\check{\Omega}_M}\\
		\leq&2\norm{\frac{1}{|D_M|}\sum_{s=0}^{\infty}\omega\Big(\frac{s}{b_M}\Big)\sum_{i\in D_M}\sum_{j\in D_M, s\leq \rho(i,j)< s+1}\mathbbm{E}\big[q(X_j,\theta^*_M)|\bm{z}\big]\tilde{q}(X_i,\theta^*_M)}\\
		=&o_p(1).
	\end{aligned}
\end{equation}
Let $B_i=\sum_{s=0}^{\infty}\omega\big(\frac{s}{b_M}\big)\sum_{j\in D_M, s\leq \rho(i,j)< s+1}\mathbbm{E}\big[q(X_j,\theta^*_M)|\bm{z}\big]$.
\begin{align*}
	&\norm{\frac{1}{|D_M|}\sum_{s=0}^{\infty}\omega\Big(\frac{s}{b_M}\Big)\sum_{i\in D_M}\sum_{j\in D_M, s\leq \rho(i,j)< s+1}\mathbbm{E}\big[q(X_j,\theta^*_M)|\bm{z}\big]\tilde{q}(X_i,\theta^*_M)}_1\\
	\leq &\norm{\frac{1}{|D_M|}\sum_{i\in D_M}\tilde{q}(X_i,\theta^*_M)B_i}_2\\
	\leq & \Big[\frac{1}{|D_M|^2}\sum_{i\in D_M}\mathbbm{E}\big(\tilde{q}(X_i,\theta^*_M)^2|\bm{z}\big)B_i^2+\frac{1}{|D_M|^2}\sum_{i\in D_M}\sum_{j\in D_M,j\neq i}\mathbbm{E}\big(\tilde{q}(X_i,\theta^*_M)\tilde{q}(X_j,\theta^*_M)|\bm{z}\big)B_iB_j\Big]^{1/2}\\
	\leq &\Big[\frac{C_1}{|D_M|}b_M^{2d}+\frac{C_2}{|D_M|^2}\sum_{i\in D_M}\sum_{s=1}^\infty\sum_{j\in D_M,s\leq \rho(i,j)<s+1}\tilde{\kappa}_{M,s}B_iB_j\Big]^{1/2}\\
	\leq &\Big[o(1)+\frac{C_2}{|D_M|}\sum_{s=1}^\infty s^{d-1}b_M^{2d}\tilde{\kappa}_{M,s}\Big]^{1/2}=o(1).
	\tag{\stepcounter{equation}\theequation}
\end{align*}
Hence, equation (\ref{eq:last}) follows from Markov's inequality. Theorem 4.3 follows by continuity of matrix inversion and multiplication.

\section{Spillover Effect}\label{spilloversection}
In addition to the direct average treatment effect on the treated (DATT), empirical researchers might also be interested in spillover effects defined in equations (\ref{eq:spill1}) and (\ref{eq:spill0}). 
\begin{equation}\label{eq:spill1}
	\tau(1,g,g')=\frac{1}{|D_M|}\sum_{i\in D_M}\Big(\mathbbm{E}\big[\tilde{y}_{i2}(1,g)|W_i=1, G_i=g, z_i\big]-\mathbbm{E}\big[\tilde{y}_{i2}(1,g')|W_i=1, G_i=g', z_i\big]\Big)
\end{equation}
\begin{equation}\label{eq:spill0}
	\tau(0,g,g')=\frac{1}{|D_M|}\sum_{i\in D_M}\Big(\mathbbm{E}\big[\tilde{y}_{i2}(0,g)|W_i=0, G_i=g, z_i\big]-\mathbbm{E}\big[\tilde{y}_{i2}(0,g')|W_i=0, G_i=g', z_i\big]\Big)
\end{equation}
When the exposure mapping could be misspecified, we have
\begin{equation}
\label{eq2}
\begin{aligned}
\tau^*(1,g,g')=&\frac{1}{|D_M|}\sum_{i\in D_M}\Big(\mathbbm{E}\big[y_{i2}(1,\bm{W}_{-i})|W_i=1, G_i=g, z_i\big]\\
&-\mathbbm{E}\big[y_{i2}(1,\bm{W}'_{-i})|W_i=1, G_i=g', z_i\big]\Big)
\end{aligned}
\end{equation}
and
\begin{equation}
\label{eq3}
\begin{aligned}
\tau^*(0,g,g')=&\frac{1}{|D_M|}\sum_{i\in D_M}\Big(\mathbbm{E}\big[y_{i2}(0,\bm{W}_{-i})|W_i=0, G_i=g, z_i\big]\\
&-\mathbbm{E}\big[y_{i2}(0,\bm{W}'_{-i})|W_i=0, G_i=g', z_i\big]\Big).
\end{aligned}
\end{equation}

The spillover effect contrasts the expected potential outcomes between levels $g$ and $g'$ and could differ with or without direct treatment. A leading case would be setting $g'$ to 0. The identification of the spillover effect is more straightforward because potential outcomes under direct assignment and the specified exposures are observable. Nevertheless, I impose a further condition to facilitate causal interpretation of the spillover effects. 
\begin{condition}
\label{partialunconfound}
$\forall\ i\in D_M$, $y_{i2}(w_i,\bm{w}_{-i})\indep \bm{W}_{-i}|W_i, z_i$. 
\end{condition}
Since I average over all population units in $D_M$, there is no compositional change of the subpopulation at exposure levels $g$ and $g'$. If it is not feasible for every population unit to receive exposure $g$ and $g'$, one can average across the subpopulation composed of units eligible for both exposure levels, instead. On top of that, given a single unit $i\in D_M$, Condition \ref{partialunconfound} rules out heterogeneity bias across different exposure levels. Suppose the potential outcome is composed of a fixed outcome plus some measurement error; namely, $y_{i2}(w_i,\bm{w}_{-i})=\phi_{i2}(w_i,\bm{w}_{-i})+e_{i2}$. A sufficient condition for Condition \ref{partialunconfound} would be that given unit $i$'s own treatment status and neighborhood attributes, its measurement error does not depend on neighbors' treatment statuses. 

Under Condition \ref{partialunconfound}, 
\begin{align*}
\tau^*(1,g,g')=&\frac{1}{|D_M|}\sum_{i\in D_M}\Big(\mathbbm{E}\big[y_{i2}(1,\bm{W}_{-i})|W_i=1, G_i=g, z_i\big]-\mathbbm{E}\big[y_{i2}(1,\bm{W}'_{-i})|W_i=1, G_i=g', z_i\big]\Big)\\
=&\frac{1}{|D_M|}\sum_{i\in D_M}\bigg(\sum_{\bm{w}_{-i}\in\Omega}\mathbbm{E}\big[y_{i2}(1,\bm{w}_{-i})|W_i=1,\bm{W}_{-i}=\bm{w}_{-i},z_i\big]P(\bm{W}_{-i}=\bm{w}_{-i}|G_i=g,W_i=1,z_i)\\
&-\sum_{\bm{w}'_{-i}\in\Omega}\mathbbm{E}\big[y_{i2}(1,\bm{w}'_{-i})|W_i=1,\bm{W}'_{-i}=\bm{w}'_{-i},z_i\big]P(\bm{W}'_{-i}=\bm{w}'_{-i}|G_i=g',W_i=1,z_i)\bigg)\\
=&\frac{1}{|D_M|}\sum_{i\in D_M}\bigg(\sum_{\bm{w}_{-i}\in\Omega}\mathbbm{E}\big[y_{i2}(1,\bm{w}_{-i})|W_i=1,z_i\big]P(\bm{W}_{-i}=\bm{w}_{-i}|G_i=g,W_i=1,z_i)\\
&-\sum_{\bm{w}'_{-i}\in\Omega}\mathbbm{E}\big[y_{i2}(1,\bm{w}'_{-i})|W_i=1,z_i\big]P(\bm{W}'_{-i}=\bm{w}'_{-i}|G_i=g',W_i=1,z_i)\bigg),
\end{align*}
where $\Omega=\{0,1\}^{|D_M|-1}$.
As a result, the spillover effect contrasts the expected potential outcome with direct treatment but weighted by different conditional probabilities of neighbors' treatment realization at either exposure $g$ or $g'$. With correctly specified exposure mapping, 
\begin{equation}
	\begin{aligned}
	\tau(1,g,g')=&\frac{1}{|D_M|}\sum_{i\in D_M}\Big(\mathbbm{E}\big[\tilde{y}_{i2}(1,g)|W_i=1, G_i=g, z_i\big]-\mathbbm{E}\big[\tilde{y}_{i2}(1,g')|W_i=1, G_i=g', z_i\big]\Big)\\
	=&\frac{1}{|D_M|}\sum_{i\in D_M}\Big(\mathbbm{E}\big[\tilde{y}_{i2}(1,g)|W_i=1, z_i\big]-\mathbbm{E}\big[\tilde{y}_{i2}(1,g')|W_i=1, z_i\big]\Big).
	\end{aligned}
\end{equation}

Analogously, the doubly robust estimands for the spillover effects are 
\begin{equation}
\label{eq:spillover1}
\begin{aligned}
\tau(1,g,g')=&\mathbbm{E}_D\Bigg[\frac{W_i}{\eta(z_i)} \frac{\mathbbm{1}\{G_i=g\}}{\eta_{1g}(z_i)}\big(Y_{i2}-m_{2,1g}(z_i)\big)+m_{2,1g}(z_i)\\
&-\frac{W_i}{\eta(z_i)} \frac{\mathbbm{1}\{G_i=g'\}}{\eta_{1g'}(z_i)}\big(Y_{i2}-m_{2,1g'}(z_i)\big)-m_{2,1g'}(z_i)\Bigg]
\end{aligned}
\end{equation}
and 
\begin{equation}
\label{eq:spillover0}
\begin{aligned}
\tau(0,g,g')=&\mathbbm{E}_D\Bigg[\frac{1-W_i}{1-\eta(z_i)} \frac{\mathbbm{1}\{G_i=g\}}{\eta_{0g}(z_i)}\big(Y_{i2}-m_{2,0g}(z_i)\big)+m_{2,0g}(z_i)\\
&-\frac{1-W_i}{1-\eta(z_i)} \frac{\mathbbm{1}\{G_i=g'\}}{\eta_{0g'}(z_i)}\big(Y_{i2}-m_{2,0g'}(z_i)\big)-m_{2,0g'}(z_i)\Bigg].
\end{aligned}
\end{equation}
The asymptotic distribution of the spillover effect estimators can be established similarly by setting up a GMM problem. Notice that Condition 1 is not required for estimation or inference but merely for causal interpretation.

\section{Different Approaches to Dimension Reduction}\label{summary}
\cite{manski2013identification} and \cite{basse2018limitations} formally point out that there exist no consistent treatment effect estimators under arbitrary interference. It is therefore necessary to make dimension reduction assumptions about the interference structure in order to identify meaningful treatment effect parameters. There are different approaches to dimension reduction in the literature; see, for instance, \cite{auerbach2021local}, \cite{agarwal2022network}, \cite{emmenegger2022treatment}, and \cite{qu2022efficient}. In this section, I provide an overview of some of the leading approaches in the literature. I show how recent literature development relates to the general framework in the current paper. Each article referenced proposes different estimation methods for various causal effect estimands. My focus here is to compare the different approaches to modeling spillover effect.\footnote{It is not supposed to be a comprehensive survey.} 

\subsection{Partial Interference}
The most popular approach to dimension reduction of the interference structure is partial interference restricted within disjoint clusters. In \cite{qu2022efficient}, their potential outcome function is modeled as\footnote{The potential outcome is defined for a single cross section.} 
\begin{equation}\label{partial}
y_{c,i}(w_{c,i},\bm{w}_{c,(i),1},\cdots,\bm{w}_{c,(i),m})\equiv y_{c,i}(w_{c,i},g_{c,1},\cdots,g_{c,m}),
\end{equation}
where $c$ is the index of a cluster, $y_{c,i}$ and $w_{c,i}$ is the potential outcome and treatment assignment of unit $i$ in cluster $c$, and $\bm{w}_{c,(i),j}$ is the treatment assignment of unit $i$'s neighbors in the disjoint subset $j$ of cluster $c$. Units within each of the $m$ disjoint subsets are exchangeable. As a result, the impact of $\bm{w}_{c,(i),j}$ can be summarized by $g_{c,j}$, which measures the number of treated neighbors in subset $j$ of cluster $c$. Compared with the assumption of fully exchangeable neighbors in cluster $c$, the partition of $m$ subsets allows for more heterogeneity of neighbors' influence. This allows for a more flexible interference structure. 

If (\ref{partial}) is correctly specified, one can choose $K$ to be $\max_{c=1,\dots,C}\max_{i,j\in c}\rho(i,j)$. Given bounded cluster sizes, $K$ is finite. For all $s>K$ and any $i$, $y_i(\bm{W})-y_i(\bm{W}^{(i,s)})=0$. Therefore, potential outcomes in the form of (\ref{partial}) can be accommodated in the approach I take. A trickier question is how to partition the $m$ subsets within each cluster $c$. On top of that, partial interference might be too strong an assumption. If either the exchangeability or the partial interference assumption does not hold, the approach in the current paper can still identify the expected exposure effect as long as the interference from units further away is increasingly negligible. 

\subsection{Immediate Neighbors}
A slightly different approach to dimension reduction is to restrict interference within immediate neighbors. For instance, in \cite{emmenegger2022treatment}, 
the spillover function is specified as 
\begin{equation}\label{immediate}
\big(f^1(\{W_j\}_{j\in D_M,j\neq i}), \cdots, f^r(\{W_j\}_{j\in D_M,j\neq i})\big)
\end{equation}
of fixed dimensions $r$. Each such function is specified by empirical researchers and describes a one-dimensional spillover effect that unit $i$ receives from its neighbors. In Example 2.1 in \cite{emmenegger2022treatment}, the functions $f^l$ has been specified as the average number of treated neighbors of unit $i$ and the average number of treated neighbors of neighbors of $i$, respectively, for $r=2$. In this case, if one defines neighbors of $i$ as units within distance $\bar{K}$ from $i$, then the approximate neighborhood interference (ANI) assumption holds for any $s>2\bar{K}$.

Equations (\ref{partial}) and (\ref{immediate}) have recently been proposed in the literature allowing for a more flexible interference structure. The purpose of the discussion is to show that if empirical researchers assume these specifications of the spillover function are correct, they can be well accommodated in the framework of the current paper. Even if some dimension reduction assumptions fail, applied researchers are still able to identify causal estimands as long as ANI is true.
 
\subsection{Local Configuration}
A more interesting discussion is the comparison of the local configuration approach proposed by \cite{auerbach2021local} and ANI. In a spatial setting, unit $i$’s local configuration of radius $r$, denoted by $G_i^r$, refers to the units within distance $r$ of $i$ and their treatments and characteristics.\footnote{To maintain consistency, I follow the notation in \cite{auerbach2021local}. Therefore, with the abuse of notation, $G_i$ is different from the exposure mapping defined above.} Units within a local configuration remain anonymous, similar to the exchangeability assumption. ANI and the expected exposure mapping are initially proposed to allow for misspecification of the spillover function. 
The local configuration approach instead maintains correct specification of the spillover function. However, it uses local configurations of various radius $r$ to approximate the effective treatment according to the spillover function.\footnote{See \cite{manski2013identification} for the definition of ``effective treatment." The terms ``exposure mapping" from \cite{aronow2017estimating} and ``effective treatment" from \cite{manski2013identification} are used interchangeably throughout the text.} Below, I provide another interpretation of the ANI assumption. Under correct specification of the spillover function, the ANI approach is not too different from the local configuration approach. 

According to the metric definition in \cite{auerbach2021local}, for effective treatment $g$ and $\tilde{g}$, if the distance $d(g,\tilde{g})\leq \frac{1}{1+r}$ then $G_i^r=\tilde{G}_i^r$. Under Assumption 4.5 therein, 
\begin{equation}\label{local}
\big|h(g_0)-h(\tilde{g})\big|\leq \phi\big(d(g_0,\tilde{g})\big),
\end{equation}
where $\phi(x)\to 0$ as $x\to 0$, $h(g)=\mathbbm{E}[h(g,U_i)]$, and $Y_i=h(G_i,U_i)$. Therefore, we can see that (\ref{local}) goes to 0 as $r\to \infty$, which is analogous to the ANI assumption in \cite{leung2022causal}.
\begin{equation}\label{ani}
\sup_{M}\max_{i\in D_M}\mathbbm{E}\Big[\big|Y_i(\bm{W})-Y_i\big(\bm{W}^{(i,r)}\big)\big|\Big]\to 0,\text{ as } r\to\infty
\end{equation}

Examples 2.1 and 2.2 in \cite{auerbach2021local} are essentially examples of Sections C.1 and C.2, and hence I focus on their Example 2.3 -- the linear-in-means peer effects model. Assuming correct specification, 
\[
Y_i=\alpha+\delta\frac{1}{n_i}\sum_{j\in P_i}Y_j+W_i\gamma+e_i,
\]
where $P_i$ is the peer group of unit $i$ with size $n_i$. As usual, $|\delta|<1$. The reduced form of the potential outcome is solved to be 
\[
Y_i=\lim_{S\to\infty}\sum_{s=1}^Sh_s(G_i^s,U_i)=h(G_i,U_i)
\]
for some functions $h_s$ and $h$.
Hence, for $d(g,\tilde{g})\leq \frac{1}{1+r}$, 
\[
\big|h(g)-h(\tilde{g})\big|\leq C|\beta|^r \text{ for some } |\beta|<1, 
\]
which is exactly the ANI coefficient given in Proposition 1 in \cite{leung2022causal}.\footnote{I refer readers to \cite{auerbach2021local} for the introduction to notation and more detailed derivation.} 

Therefore, under a true interference structure, if one chooses a large enough $r$ neighborhood, the ANI approach can be thought of as using units with the effective treatment closest to the actual effective treatment $g$ to estimate the policy effect. 

\section{Practical Guide}\label{sec:guide}
The following steps summarize the estimation procedure of the DATT (EDATT). Spillover effects can be estimated in a similar manner. 

\begin{enumerate}
\item Collect the data $\{Y_{it},W_i,z_i\}_{i\in D_M}$ from the population.
\item Specify the exposure mapping function $G(i,\cdot)$ as a function of $\bm{W}_{-i}$.
\item Set up models for the propensity scores $P(W_i=1|z_i)$, $P(G_i=g|W_i=1, z_i)$, and $P(G_i=g|W_i=0, z_i)$. Also set up models for the conditional means of the outcomes in both time periods $\mathbbm{E}(Y_{it}|W_i=w,G_i=g,z_i)$, or the model for the conditional mean of the difference of the outcomes $\mathbbm{E}(Y_{i2}-Y_{i1}|W_i=w,G_i=g,z_i)$.
\item Combine the moment conditions based on models in Step 3 (e.g., first-order conditions from maximum likelihood estimation or linear regression) and the identification equation (\ref{eq:dr}) in the main text. 
\item Estimate the GMM model given in Step 4 and conduct spatial-correlation robust inference. The DATT (EDATT) estimate is the last element of the GMM estimates. 
\end{enumerate}

\section{Additional Simulation Results}\label{sec:addsim}
I examine the inference performance of doubly robust estimators with finite samples in this section. In the main text, Section \ref{sec:simulation} describes how the population is generated.
The standard deviation of the $\tau(1)$ estimates is summarized in the top panel of Table \ref{tab:3} below. Regression adjustment comes with the smallest standard deviation. It is more interesting to see that the standard deviation of the doubly robust estimates can be one third smaller than that of the IPW estimates. With moderate misspecification, we can still see efficiency gains from using the doubly robust estimator. 

\begin{table}[htbp]
  \centering
  \caption{Standard Deviation and Coverage of CI: $\tau(1)$}
\begin{threeparttable}
    \begin{tabular}{lcccccc}
\toprule
      & 1     & 2     & 3     & 4     & 5     & 6 \\
\midrule
&\multicolumn{6}{c}{standard deviation}\\
\cmidrule{2-7}\\
    ra    & 0.154 & 0.154 & 0.155 & 0.163 & 0.324 & 0.156 \\
    ipw\_mle   & 0.205 & 0.252 & 0.305 & 0.340 & 0.532 & 0.305 \\
    ipw\_cbps  & 0.206 & 0.252 & 0.306 & 0.341 & 0.530 & 0.306 \\
    dr\_mle & 0.173 & 0.173 & 0.194 & 0.211 & 0.570 & 0.194 \\
    dr\_cbps & 0.173 & 0.173 & 0.194 & 0.211 & 0.562 & 0.194 \\
\midrule
&\multicolumn{6}{c}{coverage rate}\\
\cmidrule{2-7}\\
    cov\_ehw   & 0.947 & 0.947 & 0.941 & 0.932 & 0.932 & 0.942 \\
    cov\_0.6   & 0.945 & 0.946 & 0.939 & 0.935 & 0.944 & 0.940 \\
    cov\_1     &  0.944 & 0.944 & 0.937 & 0.934 & 0.945 & 0.938 \\
\bottomrule
    \end{tabular}%
   \begin{tablenotes}
   \begin{footnotesize}
   \item[1] The coverage rate is based on the standard error of the doubly robust estimator with CBPS moment conditions.
\item[2] cov\_ehw stands for the coverage rate of the 95\% confidence interval based on the EHW standard error; cov\_0.6 stands for the coverage rate of the 95\% confidence interval based on the SHAC standard error with bandwidth 0.6; cov\_1 stands for the coverage rate of the 95\% confidence interval based on the SHAC standard error with bandwidth 1.    
\item[3] The confidence intervals are centered on the average of point estimates. Thus, the coverage rate simply compares the magnitudes of the standard errors without taking into account the bias of the point estimates. 
\end{footnotesize}
\end{tablenotes}
\end{threeparttable}
  \label{tab:3}%
\end{table}%

The bottom panel of Table \ref{tab:3} summarizes the coverage rate of the 95\% confidence interval based on the usual standard error of the doubly robust estimator with CBPS moment conditions. In this population generating process, the EHW standard errors work well in designs 1-3 and 6. In designs 4-5, misspecification of the linear-in-means outcome model induces more spatial correlation. As a result, the confidence interval based on the SHAC standard errors provides better coverage than that based on the EHW standard errors. 

I introduce an additional design with heterogeneous direct treatment effects. There are now 900 units in the lattice. Among them, 612 units have neighbors and are thus eligible for spillover. The individual treatment assignment probability remains as $p(z^*)=\frac{exp(0.3z+0.8z_u)}{1+exp(0.3z+0.8z_u)}$ but the second period potential outcomes are $\bm{Y}_2=2+3\bm{z}\odot\bm{W}+0.2A*\bm{Y}_2+2\bm{z}^2+\bm{e}_2$. As a result, $\tau(1)=\tau(0)=0$ for the entire population. For the subpopulation composed of units with neighbors, $\tau(1)=\tau(0)=-0.016$. Point estimates follow similar patterns as in Table \ref{tab:2} in the main text. The biases of the doubly robust estimator with CBPS moment conditions are -0.017 and -0.021 for $\tau(1)$ and $\tau(0)$, respectively. In this design, we do see (substantive) over coverage of the 95\% confidence intervals for the average direct effect and parameters in the other moment conditions in the GMM estimation.  

Table \ref{tab:append1} below summarizes results for a subset of the GMM parameters. The first five columns are coverage rates for the parameters in $q_2$, the moment condition for the propensity score for $G$. The next five columns are coverage rates for the parameters in the outcome regression moment condition in the second time period, $q_4$. The last two columns are coverage rates for the two direct effects at exposure levels one and zero. 
Because of the spatial correlation induced by spillover, the SHAC standard errors are the appropriate ones to be considered. As expected, the EHW standard errors can be a bit too small when spatial correlation is nonnegligeble. Because the usual standard errors tend to be conservative, the coverage rates of the confidence interval constructed using the SHAC standard errors with appropriate bandwidth can go above the nominal level of 0.95 with some coverage rates above 0.99.

\begin{table}[htbp]
\small
  \centering
  \caption{Coverage of 95\% Confidence Intervals}
\begin{threeparttable}
    \begin{tabular}{l|ccccc|ccccc|cc}
\toprule
    cov\_ehw  & 0.869 & 0.845 & 0.964 & 0.950 & 0.914 & 0.965 & 0.963 & 0.991 & 0.969 & 0.955 & 0.949 & 0.928 \\
    cov\_0.6 &   0.942 & 0.928 & 0.963 & 0.955 & 0.956 & 0.966 & 0.967 & 0.990 & 0.968 & 0.973 & 0.955 & 0.943 \\
    cov\_1     & 0.957 & 0.942 & 0.961 & 0.959 & 0.965 & 0.966 & 0.966 & 0.990 & 0.968 & 0.977 & 0.956 & 0.947 \\
    cov\_1.4   & 0.960 & 0.946 & 0.960 & 0.959 & 0.967 & 0.965 & 0.967 & 0.990 & 0.968 & 0.977 & 0.956 & 0.948 \\
\bottomrule
    \end{tabular}%
\begin{tablenotes}
\item[1] The results are for the doubly robust estimator with CBPS moment conditions for the propensity scores in the GMM estimation;
\item[2] In columns 1-5, the coverage rate is with regard to the parameters in the propensity score moment condition for $G$, $q_2$; in columns 6-10, the coverage rate is with regard to the parameters in the outcome regression moment condition, $q_4$; in the last two columns, the coverage rate is with regard to $\tau(1)$ and $\tau(0)$. 
\end{tablenotes}
\end{threeparttable}
  \label{tab:append1}%
\end{table}%

\section{Data Generating Process for Table \ref{tab:compare}}\label{sec:DGP}
The data generating process is largely based on that in Section \ref{sec:simulation} with $|D_M|=400$ units. I make the following modifications. 
The assignment variable $W$ is equal to one when the input value $\xi_i$ is greater than its population average $\sum_{i\in D_M}\xi_i/|D_M|$, where $\xi_M$ is a $|D_M|\times 1$ vector drawn from a multivariate normal distribution with mean zero and a variance-covariance matrix equal to 0.3 raised to the power of the distance between units. The exposure mapping is specified as $G=\mathbbm{1}\{A\bm{W}>0\}$, the same way as in Section \ref{sec:simulation}. 

In the first design, $Y_2=2+W*G+z+e_2$ with $\tau(1)=1$ and $\tau(0)=0$. In the second design, $Y_2=2+(1-W)*G+z+e_2$ with $\tau(1)=-1$ and $\tau(0)=0$. 
The potential outcome function in the first time period is always $y_1(0,\underline{0})=1+z+e_1$.
$\tau_{canonic}$ reported in Table \ref{tab:compare} is based on the TWFE estimates averaged across 1,000 replications. $\tau$ is computed as the weighted average of direct effects at different exposure levels. In the simulation, $P(G_i=1|W_i=1,z_i)-P(G_i=1|W_i=0,z_i)$ on average is 0.538, which is approximately the gap between $\tau_{canonic}$ and $\tau$ when there is spillover effect on the non-directly treated units.

\end{document}